\documentclass[10pt,journal,cspaper,compsoc]{IEEEtran}

\usepackage{graphicx}
\usepackage{amsmath,amssymb}
\usepackage{subfigure}
\usepackage{algorithmic}
\usepackage{listings}
\usepackage{multirow}
\usepackage{color}

\usepackage{lscape}

\newcommand{\tc}[0]{\textcolor{black}}

\newcommand{\squishlist}{
\begin{list}{$\bullet$} {
	\setlength{\itemsep}{0pt}
	\setlength{\parsep}{3pt}
	\setlength{\topsep}{3pt}
	\setlength{\partopsep}{0pt}
	\setlength{\leftmargin}{1.5em}
	\setlength{\labelwidth}{1em}
	\setlength{\labelsep}{0.5em}
	}
}

\newcommand{\squishend}{
	\end{list}
}

\begin{document}

\title{Leveraging Peer Centrality in the Design of Socially-Informed Peer-to-Peer Systems}

\author{Nicolas~Kourtellis 
	and~Adriana~Iamnitchi
\IEEEcompsocitemizethanks{
\IEEEcompsocthanksitem Nicolas Kourtellis and Adriana Iamnitchi are with the Department of Computer Science and Engineering, College of Engineering, University of South Florida, Tampa, FL, 33620.\protect\\
E-mail: nkourtel@mail.usf.edu and anda@cse.usf.edu.\protect\\
}
\thanks{}
}

\markboth{}%
{Kourtellis \MakeLowercase{\textit{et al.}}: Leveraging Peer Centrality in the Design of Socially-Informed Peer-to-Peer Systems}

\IEEEcompsoctitleabstractindextext{%

\begin{abstract}

Social applications mine user social graphs to improve performance in search, provide recommendations, allow resource sharing and increase data privacy.
When such applications are implemented on a peer-to-peer (P2P)
architecture, the social graph is distributed on the P2P system: the traversal of the social graph translates into a socially-informed routing in the peer-to-peer layer.

In this work we introduce the model of a projection graph that is the result of decentralizing a social graph onto a peer-to-peer network.
We focus on three social network metrics: degree, node betweenness and edge betweenness centrality and analytically formulate the relation between metrics in the social graph and in the projection graph.
Through experimental evaluation on real networks, we demonstrate that when mapping user communities of sizes up to 50-150 users on each peer, the association between the properties of the social graph and the projection graph is high, and thus the properties of the (dynamic) projection graph can be inferred from the properties of the (slower changing) social graph.
Furthermore, we demonstrate with two application scenarios \tc{on large-scale social networks} the usability of the projection graph in designing social search applications and unstructured P2P overlays.

\end{abstract}

}

\maketitle

\section{Introduction}\label{sec:into}

Socially-aware applications and services have leveraged social relationships for diverse objectives such as improving security~\cite{yu06sybilguard}, inferring trust~\cite{maniatis05lockss}, providing incentives for resource sharing~\cite{tran08friendstore}, and building overlays~\cite{popescu04turtle} for private communication. 
Online social information has been used to rank Internet search results relative to the interests of a user's neighborhood in the social network~\cite{gummadi06exploiting}, to favor socially connected users in a BitTorrent swarm~\cite{pouwelse08tribler}, and to reduce unwanted communication~\cite{mislove08ostra}.

\begin{figure}[h]
\centering
\includegraphics[scale=0.35]{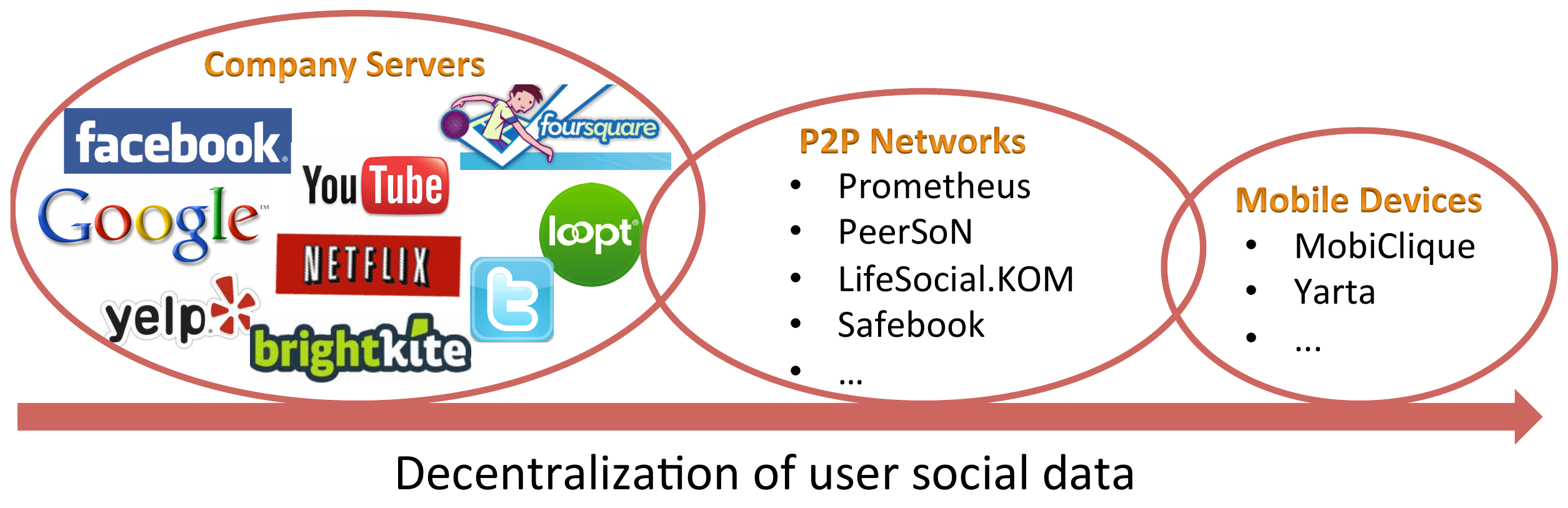}
\caption{Users' social information can be stored on a wide range of system architectures.
A node in a P2P network can hold information about a much smaller set of users than the centralized servers of a company, but a much larger set of users than a mobile device, where typically only the user-owner's social circle is known to the device.}
\label{fig:decentralizationspectrum}
\end{figure}

All such applications collect and manage social data of users in the form of a social graph ($SG$) within an application domain.
A user's social data, which consist of the user's direct relations with other users in the application, could be stored on a wide range of system architectures, as shown in Figure~\ref{fig:decentralizationspectrum}.
On one side of the spectrum, these data could be stored on centralized company servers, such as in Google and Facebook.
\tc{Centralized solutions can offer improved services and privacy protection but typically monopolize data access and use for for-profit monitoring.}
On the other side of the spectrum, they could be stored on the users' mobile devices in a fully decentralized fashion~\cite{mokhtar09middlewarepsn,pietilainen09mobiclique,sarigol10tuplespace,toninelli11yarta}.
In between, there is a wide range of distributed solutions where multiple users can have their social information stored on the same peer.
Of the many distributed architectures, a P2P architectural approach has significant benefits and has been chosen for several systems~\cite{buchegger09peerson,shakimov09visavis,cutillo09safebook,graffi11lifesocialkom,kourtellis10prometheus}.
\tc{It can provide users better control over their own data by
  obstructing the data monopoly of centralized social network providers and better service availability for social services mining the $SG$ than mobile devices.
Given such P2P solutions, the present work addresses the following question:}
\emph{How does the $SG$ topology and its decentralization affect the network properties of peers and the routing in the P2P system?}

To answer this question, we define the \emph{projection graph} ($PG$)~\cite{kourtellis11p2pcentrality} emerging from the $SG$ decentralization on a P2P system, and study its network properties.
This graph is an undirected, weighted graph whose nodes are peers contributed by users and responsible for a set of users in the $SG$, and whose \tc{$PG$ edges} connect peers whose users are socially connected.
The weights on the $PG$ edges are the number of \tc{$SG$ edges connecting} users mapped on the end peers.
We focus on three representative metrics, known in social network analysis as centrality measures:
i)~degree centrality, which shows how many peers can be contacted directly with a message broadcast,
ii)~node betweenness centrality, which quantifies the extent to which a peer controls communication between two other peers,
and iii)~edge betweenness centrality, which quantifies how much a connection between peers is utilized during communications across separate parts of the network.
Studying these centrality measures can reveal important characteristics of the system such as network hubs (using degree centrality), or on which peers an application should place data caches for reduced latency to locate data (using node betweenness), or to enhance system fault tolerance to malicious attacks by monitoring and blocking malware traffic on particular network connections (using edge betweenness).

In this paper, we investigate how the $PG$ centrality metrics correlate with the $SG$ metrics, while varying the degree of social data decentralization in the system.
A $SG$ is typically slowly changing~\cite{leskovec08evolutionOSN}: besides infrequent events such as moving to a new place or joining a new community, people rarely change their social relations.
However, the typical churn of a P2P system translates into a much more dynamic network, and thus a dynamic $PG$.
Consequently, being able to calculate (or estimate) these centrality metrics in the $PG$, independently of the rewiring of the P2P overlay but only based on the more stable $SG$, can lead to important gains \tc{in application and system design.}

First, there are direct consequences for the performance of social applications mining a $SG$ distributed on a P2P system.
For example, social search~\cite{gummadi06exploiting, yang09p2psearch} is a method of connecting users to context-relevant content made available by their friends.
A social search query follows contextually relevant \tc{$SG$ edges} over multiple social hops.
Depending on how the neighborhood of the user submitting the query is distributed on the P2P system, such a search could visit several peers, some more socially resourceful than others.
Identifying the more resourceful peers---for example, in terms of the number of social connections between users mapped on different peers (i.e., peer degree centrality)---can improve significantly the search performance~\cite{adamic01search, lin07socialunstructured}\tc{, by reducing the overall communication overhead and maximizing the success rate.}
Similarly, a socially-aware information dissemination application can target the peers through which most of the social traffic passes (i.e., high peer betweenness centrality) for fast dissemination and high \tc{system} coverage.

Second, there are significant benefits in the P2P overlay organization.
For example, in the presence of high peer churn, users could change their storage peers for better data availability.
The newly selected peers could estimate their importance in the topology based on the centrality of their users, which can be computed infrequently.
\tc{In addition, system peers storing data of highly central users could also become central and be overwhelmed by application requests.
The system could infer the appearance of such central peers based on their users' centrality and 1)~monitor the socially-routed P2P traffic through them, 2)~place data caches or replicas, and 3)~alleviate bottlenecks by remapping high betweenness users onto better provisioned peers.}

The contributions of this study are the following:
\squishlist
\item It presents a formal model for the $PG$s in P2P systems (Section~\ref{sec:model}).
\item It studies the analytical relations of the three social network metrics between users in the $SG$ and their peer in the $PG$ (Section~\ref{sec:sna-measures}).
\item It examines experimentally on real networks the association between centrality of users and peers and estimation methods for the centrality metrics of the peers from the scores of their users, while varying the degree of social data decentralization in the system(Sections~\ref{sec:pg-setup} and~\ref{sec:association-sg-pg}).
\item It demonstrates on large-scale social networks the benefits of using $PG$ properties for applications traversing a $SG$ and for organizing the overlay of an unstructured P2P system, for improved success rate and reduced overhead (Section~\ref{sec:scenarios-application-overlay}).
\item It outlines a set of lessons that connect previous work (discussed in Section~\ref{sec:rel-work}) on $SG$s and P2P systems with the $PG$ model and shows how our findings can be applied in the design of socially-aware applications and P2P systems (Section~\ref{sec:discussion}).
\squishend

\section{Projection Graph Model}\label{sec:model}

\tc{Social graphs can be distributed on P2P systems in various ways.
The approach most used so far is based on resource or system optimization objectives: for example, in distributed hash tables (DHT) the mapping of user data on peers is done randomly to optimize load balancing.
An alternative is socially-aware distribution of data onto peers.  
P2P systems such as Prometheus~\cite{kourtellis10prometheus} showed the benefits of social-based distribution of the $SG$ in terms of resilience to attacks and data access latency for social applications~\cite{kourtellis12thesis}.}

\tc{The $PG$ model covers both approaches, as its definition is independent of the mechanism used for the distribution of data. 
However, because it has been less explored, we focus on social mappings,  which also lead to more interesting correlations between $SG$s and $PG$s.
The following scenarios provide the intuition for the social-based $PG$ and motivate the study of its network properties.}

\subsection{Motivating Scenarios}\label{sec:motivating-scenarios}

\textbf{\emph{Civilian Networking in Large-Scale Disaster:}}
After a natural disaster of large proportions, most of the communications and IT infrastructure is destroyed.
Survivors cluster around small communities such as local organizations and community centers in villages, etc.
To help with the emergency information dissemination as well as the organization of search-and-rescue operations, the authorities equip these communities with some basic IT and communication equipment (e.g., commodity servers, GSM/Wifi networks).
Community members input on these machines (peers) their health status, as well as the status (e.g., alive, injured, deceased or unknown) of their close family and friends living with them or in close distance in the area of the disaster.
This information represents the \tc{$SG$ edges} that connect users in the same geo-localized community.
Individuals may also input information about friends located in other communities along with any useful detail regarding their status (e.g., last time seen, etc).
The community machines are connected with each other over wired and wireless networks and form a rudimentary P2P network.
In this way, a basic social network of civilians is formed and their social information is stored on these machines in a decentralized fashion.

\textbf{\emph{Player Networking in Online Games:}}
Online gaming platforms (such as Steam~\cite{steamcommunity}) allow players from around the world to run servers (peers) and host multi-player gaming sessions.
Players typically choose a server as their favorite, due to low network delays or the player community on the server.
Many of the in-game and social interactions between players are stored on the server as meta-game social edges.
Occasionally, gamers play on different servers, for example when their ``home'' server is offline or overloaded.
In this way, they also form social edges with players from other communities.

\tc{In both scenarios, a $PG$ emerges when a $SG$ naturally partitioned into social communities is distributed across the P2P network on community-owned peers.
The assumption is that peers are user-contributed and serve the social community of which the user is a member. 
The scenarios above have embedded two implicit incentives for community contributions. 
Other application-specific scenarios are obvious and along the line of P2P systems.}

\subsection{Formal Model Definition}

\begin{figure}[h]
\centering
	\includegraphics[scale=0.45]{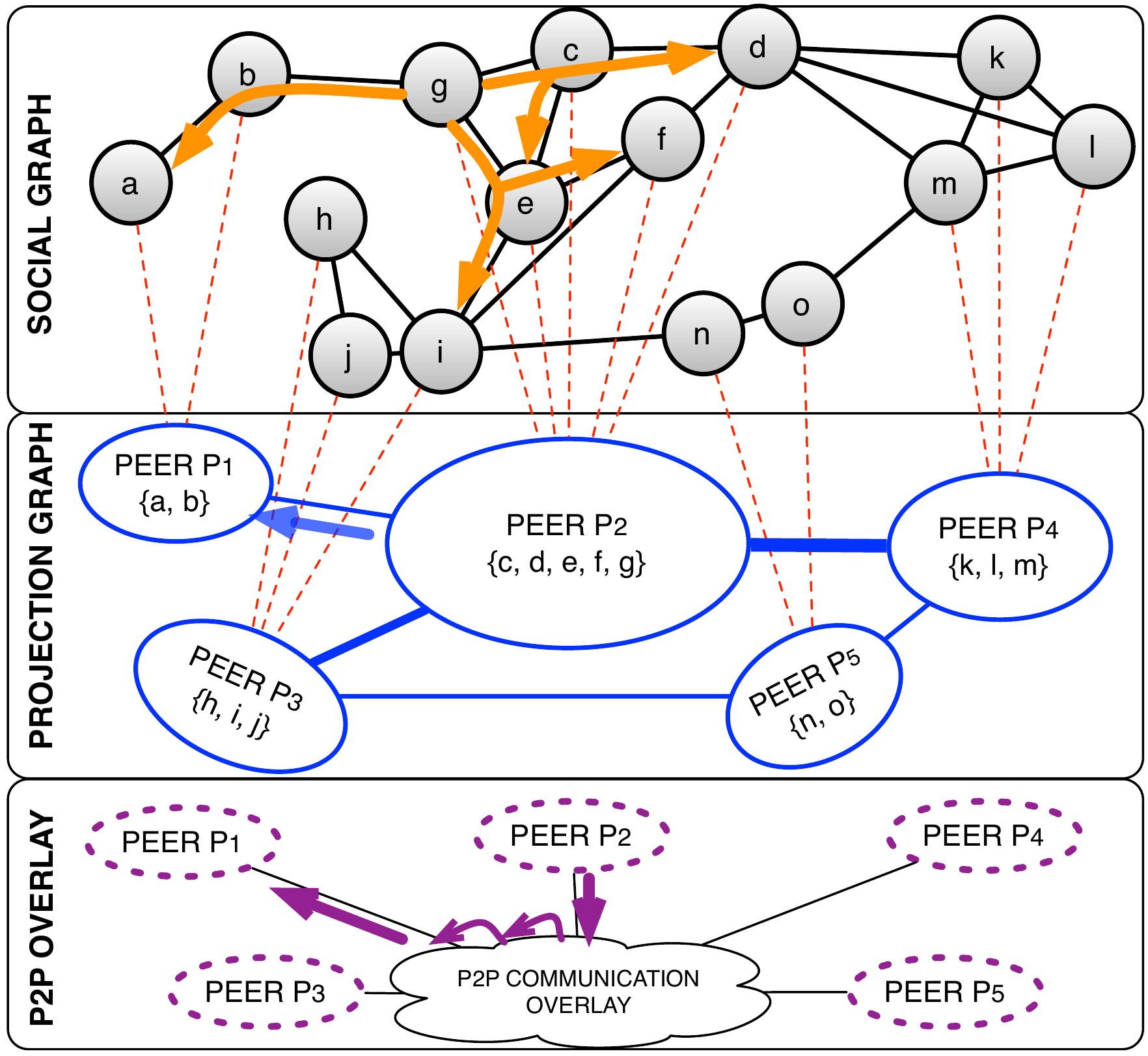}
	\caption{An example of a $SG$ distributed on a set of peers organized in a P2P overlay.
	\tc{Users $a$--$o$ are shown in small grey circles, peers $P_1$--$P_5$ are shown in large blue circles in the $PG$ and purple dashed circles declare the peers organized in a P2P overlay.}
	Users are connected with \tc{$SG$ edges} illustrated with black lines, blue lines correspond to $PG$ edges with different weights (declared by their width) and red dashed lines correspond to mappings of users onto peers storing their data.}
      \label{fig:socialgraph-p2p-new}
\end{figure}

We consider a $SG$ as an undirected and unweighted graph $SG=(V_S,E_S)$, with set of users $V_S$ and set of edges $E_S\subseteq V_S \times V_S$, representing the ties between users (top layer of Figure~\ref{fig:socialgraph-p2p-new}).
\tc{A $SG$ edge between users $u$ and $v$ $(u,v)$ is undirected and unweighted, i.e., with $w(u,v)$=$1$.}

The $PG$ in a P2P system emerges when the $SG$ is distributed on the P2P network (middle layer of Figure~\ref{fig:socialgraph-p2p-new}).
The $PG$ is an undirected, weighted graph whose nodes are peers responsible for a set of users in the $SG$ and whose edges represent the social ties between the users mapped on different peers.
We refer to a user $u$ as ``mapped" on a particular peer when the peer stores $u$'s \emph{social data} (the set of all $SG$ edges originating from $u$).

Formally, a $PG$ is represented by $PG$=$(V_P,E_P)$, with set of peers $V_P$ in the P2P system.
For each peer $P_i$$\in$$V_P$,  $\Gamma_i$  is the set of users mapped on $P_i$.
$E_P\subseteq V_P \times V_P$ is the set of edges in $PG$.
A $PG$ edge between $P_i$ and $P_j$, where $i \neq j$, is formally defined as follows:
\tc{\[ (P_i, P_j) \in E_P \text{ iff } \exists u \in \Gamma_i, \exists v \in \Gamma_j, i \neq j \text{ s.t. } (u,v) \in E_S \]}
\tc{The set of $SG$ edges $\Theta_{ij}$ that connect the users mapped on peer $P_i$ with the users mapped on peer $P_j$, is formally defined as,
\[ \Theta_{ij} = \{ (u,v) \in E_S | u \in \Gamma_i, v \in \Gamma_j, i \neq j \} \]
The weight of a \tc{$PG$ edge} between $P_i$ and $P_j$ is given by the cardinality of the set $\Theta_{ij}$, and is denoted by $w(P_i,P_j)=|\Theta_{ij}|$, with $w(P_i,P_i)=0$ by definition.}

Figure~\ref{fig:socialgraph-p2p-new} presents a scenario in which users $a$--$o$ store their data on peers $P_1$--$P_5$, and each peer has access to all its users' data.
\tc{The $PG$ edge $(P_2,P_4)$ has weight $w(P_2,P_4)$=$3$ given by $SG$ edges $(d,k)$, $(d,l)$ and $(d,m)$.}

\tc{In this model, a userÕs social data is stored on \emph{one peer}, and each peer stores \emph{at least} one user's social data.
Future work could incorporate data replication on multiple peers, by splitting the weight of a SG edge between its replicas or allowing peers to establish PG edges based on user-defined replica priorities.}
Each peer maintains the union of social data of the users mapped on it.
Depending on the social relationship of these users, this union can be anywhere from a disjoint set of \tc{$SG$ edges}, as proposed in~\cite{buchegger09peerson,shakimov09visavis,cutillo09safebook}, to a connected subgraph, as proposed in~\cite{kourtellis10prometheus}.

The $PG$ is independent from the P2P overlay, as explained in the following example. 
Assume an application wants to find the users in the $2$-hop social neighborhood of user \tc{$g$} (i.e., friends and friends of friends).
\tc{The application can search for these users by traversing the graph over the social ties user $g$ has with the rest of the users (black arrows on $SG$ edges}, Figure~\ref{fig:socialgraph-p2p-new}, top layer).
Since the $SG$ is distributed on top of a P2P network, these requests will be routed from peer to peer in a manner informed by the $SG$ topology.
Therefore, the $SG$ traversal dictates peer \tc{$P_2$} sending a message to peer \tc{$P_1$} (blue arrow in $PG$, Figure~\ref{fig:socialgraph-p2p-new}, middle layer) to request information regarding the $1$-hop connections of user \tc{$b$}.
This application request might translate into multiple routing hops between \tc{peers in the P2P overlay (e.g., DHT)} before the destination peer is located and the request is delivered (purple arrows in P2P overlay, Figure~\ref{fig:socialgraph-p2p-new}, bottom layer).
We call such systems \emph{socially-informed} because the communication pattern between peers is determined by the $SG$ topology \tc{and its projection on peers, and can be seen independently of the P2P overlay organization.}

\section{Related Work}\label{sec:rel-work}

The management of social data in a P2P architecture has been addressed in systems such as PeerSoN~\cite{buchegger09peerson}, Vis-{\`a}-Vis~\cite{shakimov09visavis}, Safebook~\cite{cutillo09safebook}, LifeSocial.KOM~\cite{graffi11lifesocialkom} and Prometheus~\cite{kourtellis10prometheus}.
In some cases (PeerSoN, Vis-{\`a}-Vis, Safebook, LifeSocial.KOM), the information of a user is isolated from other users, and peers access them individually.
Thus, the $SG$ is fragmented into 1-hop neighborhoods, one for each user, and distributed across all peers, with potentially multiple fragments stored on the same peer.
In contrast, in Prometheus~\cite{kourtellis10prometheus}, a peer can mine the collection of social data entrusted to it by a group of (possibly socially connected) users. 
In all these systems, regardless of the way peers are organized in the P2P architecture (e.g., in a structured or unstructured overlay), the $PG$ model can be applied for studying and improving system and application routing.

Other systems directly reflect the topology of the $SG$ of their users.
Turtle~\cite{popescu04turtle} uses trust relationships between users to build overlays for private communication and anonymity preservation.
F2F~\cite{li06f2f} uses social incentives to find reliable storage nodes in a P2P storage system.
Sprout~\cite{marti06sprout} enhances the routing tables of a Chord DHT with additional trusted social links of online friends, to improve query results and reduce delays.
Tribler~\cite{pouwelse08tribler} allows socially connected users that participate in the same BitTorrent swarms to favor each other in content discovery, recommendation and file downloading.

In other studies, such as~\cite{yang09p2psearch}, peers are organized into social P2P networks based on similar preferences, interests or knowledge of their users, to improve search by utilizing peers trusted or relevant to the search.
Similarly, in~\cite{lin07socialunstructured} a social-based overlay for unstructured P2P networks is outlined, that enables peers to find and establish ties with other peers if their owners have common interest in specific types of content, thus improving search and reducing overlay construction overhead.
In~\cite{wang08sop2psn}, P2P social networks self-organize based on the concept of distributed neuron-like agents and search stimulus between peers, to facilitate improved resource sharing and search.
In such systems, the peers form edges over similar preferences of their owners or search requests (i.e, $PG$ edges).
Thus, they implicitly use the $PG$ model to organize peers into a P2P social network.

\tc{Relevant to our work is the notion of the \emph{group-reduced graph}~\cite{everett99centrality}, where a group of users is replaced by a single ``super" vertex (similar to the peer in the $PG$ model).
However, in the $PG$ model:
1)~all users must be mapped to groups/peers, while the group reduced graph has both a super vertex and regular users as nodes;
2)~a peer is consequently connected only to other peers (and not users); and
3)~PG edges are weighted, while there is no concept of edge weight in the group-reduced model.
Moreover, the authors of the group-reduced graph model express a reservation related to the applicability of their model: from a sociological point of view it is difficult to justify the removal of $SG$ edges between users within a social group.
The $PG$, however, materializes in the technical space, and thus the relationships between users are irrelevant within the peer they are mapped on.}

Studies such as~\cite{everett99centrality} and~\cite{kolaczyk09co-betweenness} analytically discuss the betweenness centrality of a group of nodes by computing shortest paths between nodes outside the group, that pass through at least one node in the group~\cite{everett99centrality} or all the nodes in the group~\cite{kolaczyk09co-betweenness}.
Similarly, we study the betweenness centrality of peers representing groups of users.
However, we assume that \emph{all} users are mapped on peers (groups) and compute the peer betweenness centrality based on shortest paths between users mapped on different peers only.

\section{Social Network Centrality Metrics}\label{sec:sna-measures}

This section formally defines the degree, node betweenness and edge betweenness centrality of a $SG$ and its corresponding $PG$.
We study the connection between $SG$ and $PG$ measures and formulate research questions that we answer experimentally.
In the following, we assume that multiple users can be mapped on the same peer and a user can be mapped only on one peer.

\subsection{Degree Centrality}

The degree centrality~\cite{freeman79centrality} \tc{$C_D(u)$ of graph node $u$ is the number of edges $u$ has with other nodes}.
The degree centrality of user $u$ mapped on peer $P_i$ can be expressed as the \tc{$SG$ edges} of $u$ with users mapped on $P_i$, and the \tc{$SG$ edges with users mapped on peers other than $P_i$}:
\tc{\begin{equation}\label{eq:du}
C_D(u) =
\sum_{\substack{
v \neq u \in \Gamma_i
}} w(u,v)
+
\sum_{\substack{
v \in \Gamma_j,\\
P_j \neq P_i \in V_P
}} w(u,v),
\forall u \in \Gamma_i
\end{equation}}

The degree centrality of $P_i$ is a function of the sum of degree centralities of users mapped on $P_i$, the number of \tc{$SG$ edges} between users mapped on $P_i$ and the number of \tc{$SG$ edges} between users on $P_i$ and $P_j,\forall P_j \neq P_i \in V_P$:
\tc{\begin{equation*}
C_D(P_i) =
\sum_{
u \in \Gamma_i
} C_D(u) -
\sum_{ \substack{
u \neq v \in \Gamma_i
}}
w(u,v) -
\end{equation*}
\begin{equation}\label{eq:dp}
\sum_{
P_j \neq P_i \in V_P
}
(
\sum_{ \substack{
u \in \Gamma_i\\
v \in \Gamma_j
}}
w(u,v)-1)
\end{equation}}

\begin{figure}[h]
  \centering
	\includegraphics[scale=0.5]{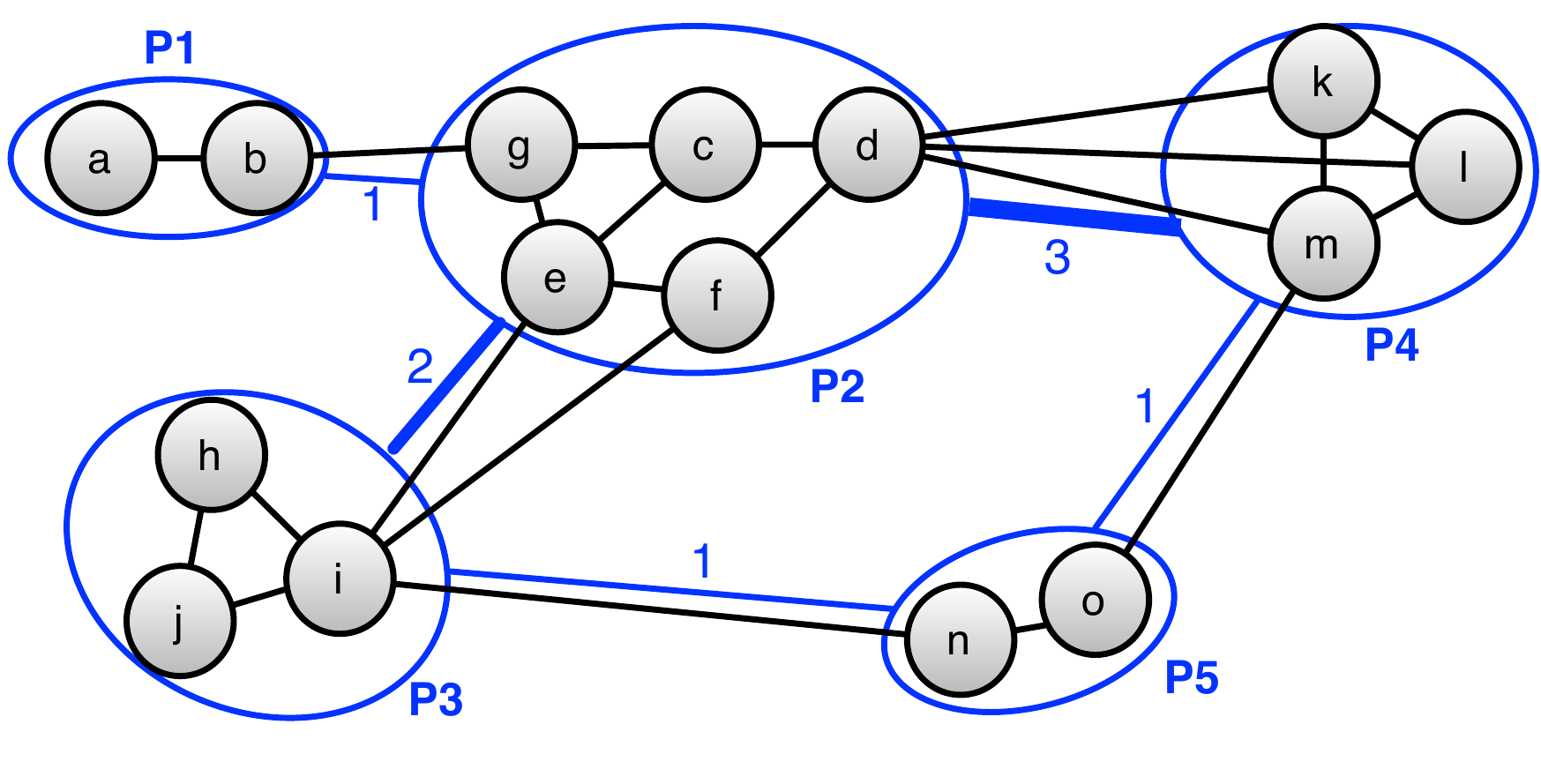}
      \caption{Example formation of a projection graph by mapping the social graph communities of users (black circles) and their social edges (black lines) onto peers (blue circles) connected over weighted peer-to-peer edges (blue lines).}\label{fig:pg-example-analysis}
	\vspace{-1mm}
\end{figure}

For example, peer $P_2$ in Figure~\ref{fig:pg-example-analysis} has a degree centrality $C_D(P_2)=3$.

Equation~\ref{eq:dp} allows us to analytically calculate the exact peer degree centrality if the peer can access its users' social connections and infer its $PG$ edges with other peers.
However, it is difficult to determine the exact degree centrality of a peer when it is granted access to view only a user's degree centrality score but not the user's neighbors and to which peers they are mapped.
Thus, a research question is:\\
\emph{\textbf{Question 1:} Can a peer estimate its $PG$ degree centrality based only on the $SG$ degree centrality score of its users?
}\label{eq:dp-question}

\subsection{Node Betweenness Centrality}

Betweenness centrality~\cite{freeman77betweenness} $C_{NB}(u)$ of a user $u$$\in$$V_S$ is the sum of fractions of shortest paths between users $s$ and $t$ that pass through user $u$, denoted by $\sigma(s,t|u)$, over all the shortest paths between the two users, $\sigma(s,t)$:
\begin{equation}\label{eq:nbu}
C_{NB}(u) =
\sum_{
s \neq t \in V_S}
\frac{\sigma(s,t|u)}{\sigma(s,t)}
\end{equation}

Betweenness centrality $C_{NB}(P_i)$ of $P_i$$\in$$V_P$ is the sum of fractions of \textit{weighted} shortest paths between $P_j$ and $P_k$ that pass through $P_i$, denoted by $\lambda(P_j,P_k|P_i)$, over all the \textit{weighted} shortest paths between the two peers, $\lambda(P_j,P_k)$:
\begin{equation}\label{eq:nbp}
C_{NB}(P_i) =
\sum_{
P_j \neq P_k \in V_P}
\frac{\lambda(P_j,P_k|P_i)}{\lambda(P_j,P_k)}
\end{equation}

\begin{figure}[h]
  \centering
	\includegraphics[scale=0.5]{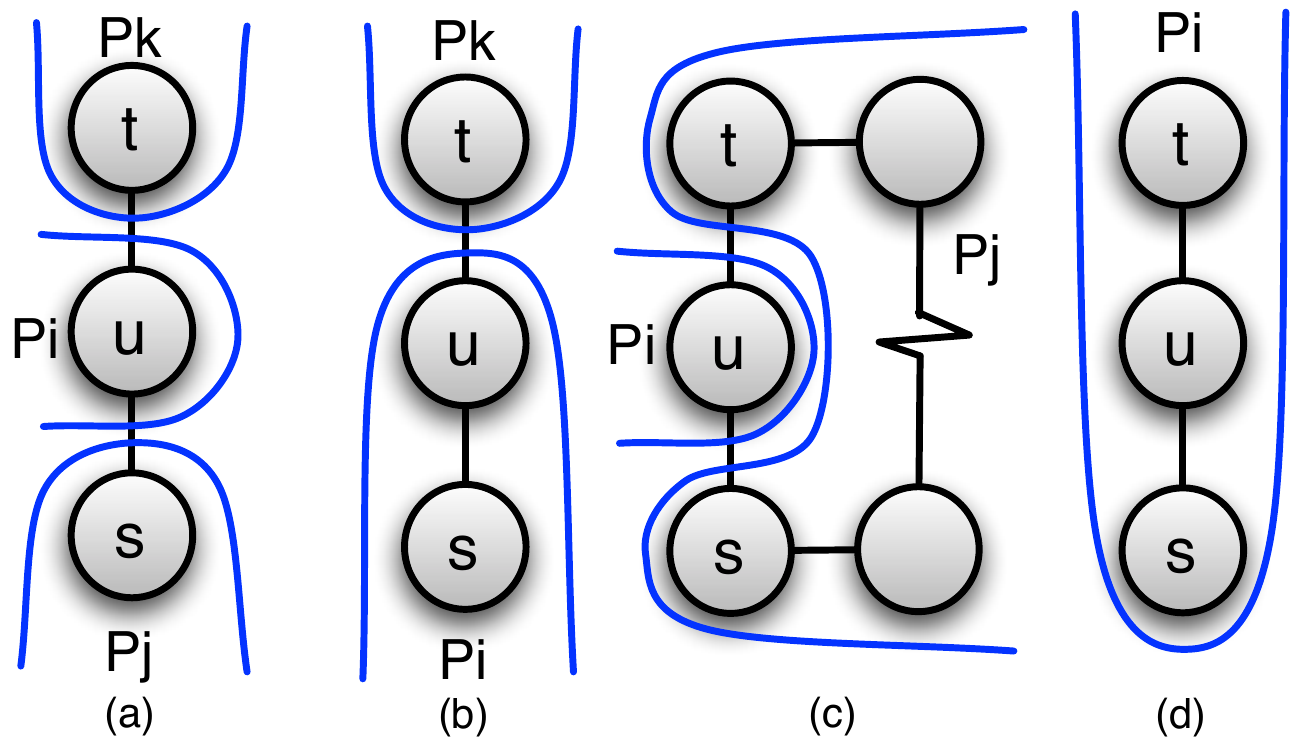}
      \caption{The four categories of shortest paths between $s$ and $t$ through $u$, when users are mapped on peers.}\label{fig:nodebetweenness-analysis}
	\vspace{-1mm}
\end{figure}

When users are mapped on peers, their shortest paths can be expressed as a combination of four basic categories, as illustrated in Figure~\ref{fig:nodebetweenness-analysis}.
The first category reflects the shortest paths between $s$ and $t$ that pass through $u$ and each user is mapped on a different peer.
The second category reflects the shortest paths between $s$ and $t$, when one of them is mapped on the same peer as $u$.
The third category reflects the shortest paths between $s$ and $t$ when they are mapped on the same peer $P_j$, but different from $u$.
The forth category reflects the case that all three users are mapped on the same peer $P_i$.

Thus, we can express the betweenness centrality of user \tc{$u$$\in$$\Gamma_i$} as a combination of these main categories of shortest paths, as follows:
\tc{\begin{equation*}
C_{NB}(u) =
\sum_{\substack {
P_j \neq P_k \in V_P
}
}
\Big(
\sum_{\substack {
s \in \Gamma_j\\
t \in \Gamma_k
}}
\frac{\sigma(s,t|u)}{\sigma(s,t)} +
\end{equation*}
\begin{equation}\label{eq:nbu-ana}
\sum_{\substack {
s \in \Gamma_i\\
t \in \Gamma_k
}}
\frac{\sigma(s,t|u)}{\sigma(s,t)} +
\sum_{\substack {
s \in \Gamma_j\\
t \in \Gamma_j
}}
\frac{\sigma(s,t|u)}{\sigma(s,t)} +
\sum_{\substack {
s \in \Gamma_i\\
t \in \Gamma_i
}}
\frac{\sigma(s,t|u)}{\sigma(s,t)}
\Big)
\end{equation}}

As demonstrated in eq.~\ref{eq:nbu-ana}, it is difficult to analytically determine the betweenness centrality of a peer with respect to the centrality of its users due to the various types of shortest paths in which users participate.
Also, the peer might not be granted access to traverse the P2P topology and calculate its exact betweenness centrality in the $PG$, for example due to user access policies on other peers or unavailability of peers.
Assuming a peer is granted access to its users' betweenness centrality scores, a research question is:\\
\emph{\textbf{Question 2:} Can a peer estimate its $PG$ node betweenness centrality based only on the $SG$ node betweenness centrality score of its users?
}\label{eq:nbp-question}

\subsection{Edge Betweenness Centrality}

Betweenness centrality~\cite{girvan02community} $C_{EB}(e)$ of \tc{a $SG$ edge} $e$$\in$$E_S$ is the sum of fractions of shortest paths between $s$ and $t$ that contain $e$, denoted by $\sigma(s,t|e)$, over all the shortest paths between the two users, $\sigma(s,t)$:
\begin{equation}\label{eq:ebu}
C_{EB}(e) =
\sum_{
s\neq t \in V_S
}
\frac{\sigma(s,t|e)}{\sigma(s,t)}
\end{equation}

Betweenness centrality $C_{EB}(E)$ of a $PG$ edge $E$$\in$$E_P$ is the sum of fractions of \textit{weighted} shortest paths between $P_i$ and $P_j$ that contain $E$, denoted by $\lambda(P_i,P_j|E)$, over all \textit{weighted} shortest paths between the two peers, $\lambda(P_i,P_j)$:
\begin{equation}\label{eq:ebp}
C_{EB}(E) =
\sum_{
P_i \neq P_j \in V_P
}
\frac{\lambda(P_i,P_j|E)}{\lambda(P_i,P_j)}
\end{equation}

\begin{figure}[h]
  \centering
	\includegraphics[scale=0.45]{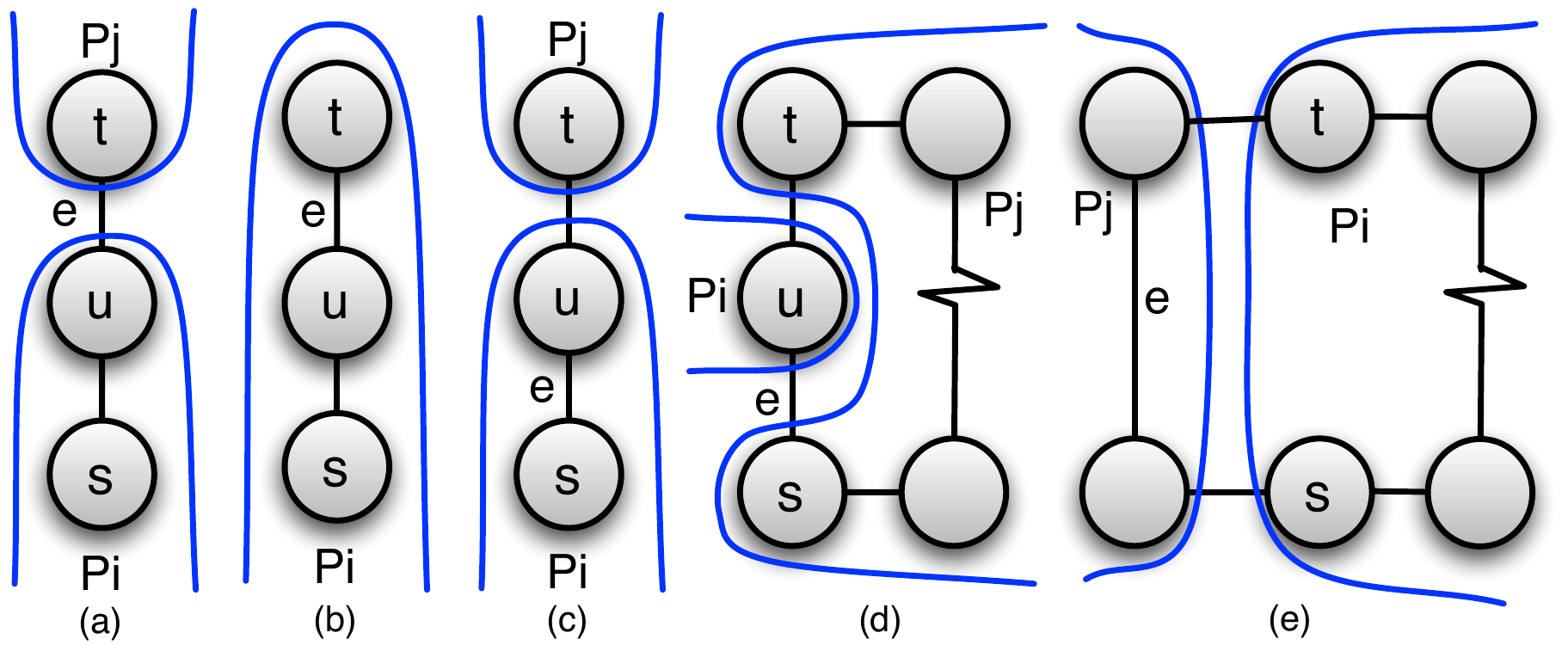}
      \caption{The five categories of shortest paths between $s$ and $t$ through $e$, when users are mapped on peers.}\label{fig:edgebetweenness-analysis}
\end{figure}

As with the node betweenness, the shortest paths between users that contain the $SG$ edge $e$ can be divided into five categories, as shown in Figure~\ref{fig:edgebetweenness-analysis}.

Based on the intuition of Figure~\ref{fig:edgebetweenness-analysis}, we can express the betweenness centrality of \tc{a $SG$ edge} $e$ as follows:
\tc{\begin{equation*}
C_{EB}(e) =
\sum_{\substack {
P_i \neq P_j \in V_P\\
}
}
\Big(
\sum_{\substack {
s \in \Gamma_i\\
t \in \Gamma_j\\
e \in \Theta_{ij}
}}
\frac{\sigma(s,t|e)}{\sigma(s,t)} +
\sum_{\substack {
s \in \Gamma_i\\
t \in \Gamma_i\\
e \in \Theta_{ii}
}}
\frac{\sigma(s,t|e)}{\sigma(s,t)} +
\end{equation*}
\begin{equation}\label{eq:ebu-ana}
\sum_{\substack {
s \in \Gamma_i\\
t \in \Gamma_j\\
e \in \Theta_{ii}
}}
\frac{\sigma(s,t|e)}{\sigma(s,t)} +
\sum_{\substack {
s \in \Gamma_j\\
t \in \Gamma_j\\
e \in \Theta_{ij}
}}
\frac{\sigma(s,t|e)}{\sigma(s,t)} +
\sum_{\substack {
s \in \Gamma_i\\
t \in \Gamma_i\\
e \in \Theta_{jj}
}}
\frac{\sigma(s,t|e)}{\sigma(s,t)}
\Big)
\end{equation}}

Using similar argumentation with the node betweenness centrality, a research question is:\\
\emph{\textbf{Question 3:} Can we estimate the edge betweenness centrality of a $PG$ edge based only on the edge betweenness centrality scores of its $SG$ edges?
}\label{eq:ebp-question}

\section{Projection Graphs From Real Networks}\label{sec:pg-setup}

In order to answer the questions stated above, we used five real networks from application domains where the $PG$ is applicable, as proposed in Section~\ref{sec:model} and \tc{constructed hypothetical $PG$ topologies in which we varied the number of users per peer.}
Table~\ref{tab:network-info} presents a summary of these networks:

\textbf{\textit{Email-Enron}}: an email communication network generated from emails sent within the Enron company.
Vertices are email addresses in the data set and undirected edges represent at least one email exchange between the end vertices.\\
\textbf{\textit{P2P-Gnutella04 and P2P-Gnutella31}}: two time-snapshots of the Gnutella peer-to-peer file sharing network on August 4th and August 31st, 2002. Nodes represent hosts in the Gnutella network topology and edges represent connections between the Gnutella hosts.
\textbf{\textit{Soc-Epinions1}}: a who-trust-whom online social network of the general consumer review site ``Epinions.com".
Nodes represent the members and edges represent the trust relationships among members.\\
\textbf{\textit{Soc-Slashdot0922}}: a network containing friend/foe links between users of "Slashdot.org", a user-contributed, technology-related news website.
Users can tag each other as friends or foes using the ``Slashdot Zoo'' feature.

These networks cover diverse domains, such as file sharing (gnutella) (from~\cite{ripeanu02gnutella}), email communications of company employees (enron), trust on consumer reviews (epinions) and friendships in a news website (slashdot) (from~\cite{lescovec11snap}) and have sizes between $10K$ and $100K$ nodes.
Even though the gnutella networks are not social networks like the other three, we use them because they exhibit social properties~\cite{ripeanu02gnutella} and we have two instances of different sizes, enabling us to study the variation of the social network metrics with the network size.
We consider all networks undirected and unweighted and used only the largest connected component (LCC) from each graph to ensure reachability between all pairs of users and peers.

\begin{table}[h]
\begin{center}
\caption{\small Summary information of the real networks used.}
\label{tab:network-info}
\begin{tabular}{|l|c|c|} \hline
Network (abbreviation)		&	Num. of Users		&	Num. of Edges	\\
						&	LCC (original)		&	LCC (original)		\\ \hline \hline
P2P-Gnutella04 (gnutella04)	&	10876 (10876)		&	39994 (39994)	\\ \hline
Email-Enron (enron)			&	33696 (36692)		&	180811 (183831)	\\ \hline
P2P-Gnutella31 (gnutella31)	&	62561 (62586)		&	147878 (147892)	\\ \hline
Soc-Epinions1	(epinions)		&	75877 (75879)		&	405739 (405740)	\\ \hline
Soc-Slashdot0922 (slashdot)	&	82168 (82168)		&	504230 (504230)	\\ \hline
\end{tabular}
\end{center}
\end{table}

\tc{To study the properties of $PG$s in such networks, we need to map users onto peers.
Since such mappings are not readily available, we decided to identify social communities on each $SG$ and then map each community to a peer.}
\tc{$SG$ edges} between communities were transformed into weighted $PG$ edges.
Communities were identified using a modified algorithm of the Louvain method~\cite{louvain08fast} for fast community detection in large networks.
This method first splits users into very small communities, and then iteratively reassigns users to other communities and merges them in order to improve the overall modularity score.

The modularity $Q$~\cite{newman04modularity} of a partition in a graph measures the density of the links inside communities as compared to links between communities, and is defined as follows:
\begin{equation*}
Q = \frac{1}{2m} \sum_{i,j} \left[ A_{ij} - \frac{k_i k_j}{2m} \right] \delta(c_i,c_j),
\end{equation*}
where $A_{ij}$ represents the weight of the edge between users $i$ and $j$, $k_i = \sum_j A_{ij}$ is the sum of the weights of the edges attached to vertex $i$, $c_i$ is the community to which vertex $i$ is assigned, the $\delta$-function $\delta (u,v)$ is 1 if $u=v$ and 0 otherwise and $m=\frac {1}{2} \sum_{ij} A_{ij}$.
The value of the modularity lies in the range $[-1, 1]$.
It is positive if the number of edges within groups exceeds the number expected on the basis of chance.
The Louvain method detects communities very fast, even for graph sizes in the order of millions of users, with a very wide range of community sizes, i.e., a lot of small groups of $2$--$10$ users, as well as very large groups in the order of 1000s users.
The largest community usually represents the \emph{core} of the network whereas the smallest ones, loosely connected to the core, reflect the \emph{whiskers} of the network~\cite{leskovec08community}.

\tc{Since we consider that a community is mapped on a user-contributed peer, it would be unrealistic to map a very large community on one peer.
Thus, we considered the communities exceeding a \textit{max-size} as individual subgraphs and recursively applied the Louvain method on them.
We call this technique ``Recursive-Louvain".
Because smaller values for \emph{max-size} dramatically increase the computation time to split the large subgraphs, we tested the Recursive-Louvain technique with selected values for the \emph{max-size}=$10$, $100$, $500$ and $1000$.
We used \emph{max-size}=$100$ as it offered the smallest standard deviation of community sizes among the tested values.
The value of \emph{max-size}=$100$ matches the findings in~\cite{leskovec08community} where the best communities with respect to \emph{conductance} were relatively small, with sizes up to $100$ users per community.}
We compare in Table~\ref{tab:louvain} the summary statistics of the formed communities with the two methods (Louvain and Recursive-Louvain) for \emph{max-size}=$100$.

\tc{Using the Recursive-Louvain method we successfully split most large communities into smaller ones ($4$ to $50$ times smaller).
Since the Louvain method tries to optimize the modularity, some communities remained larger than the \textit{max-size} applied, but within more realistic levels.
More importantly for this study, the overall variability of community size dropped (st. dev. is $6$ to $30$ times smaller). 
Thus, the majority of communities have a size close to the average, which is desirable for our experiments when varying the average community size.}

\begin{table}[htbp]
\begin{center}
\caption{Summary statistics for communities identified with Louvain ($L$) and Recursive-Louvain ($RL$) methods}
\label{tab:louvain}
\begin{tabular}{|l|l|l|l|l|}\hline
Social		&	Num. of		&						&	Standard 		&	Min/Max		\\
Network		&	Comm.		&	$\overline{|\Gamma_i|}$	&	Deviation		&				\\
			&	L / RL		&	L / RL				&	L / RL		&	L (RL Max)	\\ \hline \hline
gnutella04	&	2384/3013	&	4.0/3.6				&	23.0/3.5		&	2/1299 (89)	\\ \hline
enron		&	2434/4303	&	11.9/7.6				&	139.1/15.7	&	2/4845 (1204)	\\ \hline
gnutella31	&	13425/14385	&	4.4/4.3				&	3.0/2.8		&	2/3594 (97)	\\ \hline
epinions		&	8481/16404	&	7.1/4.6				&	196.4/7.9		&	2/15770 (484)	\\ \hline
slashdot		&	6879/18846	&	9.5/4.3				&	225.2/6.9		&	2/17012 (358)	\\ \hline
\end{tabular}
\end{center}
\end{table}

In our study, we investigate how the average number of users mapped per peer affects the estimation of the three social network measures, while using only local information on peers and $PG$ edges.
Thus, we vary the average size of communities mapped on peers.
Using the Recursive-Louvain method we identified a set of communities with fairly small average size (about $4$--$5$ users), which were used as a baseline for our experimentation with increasing average size of communities.
To produce communities with increasing number of users, we incrementally merged the smallest, socially-connected communities, until we reached the desired average number of users per community (and thus peer) in the range of $10, 20, ..., 1000$ users/peer.
This merging process finds support from~\cite{leskovec08community} where it is suggested that small communities can be combined into meaningful larger ones.

\begin{figure*}[htbp]
  \centering
	\includegraphics[scale=0.70]{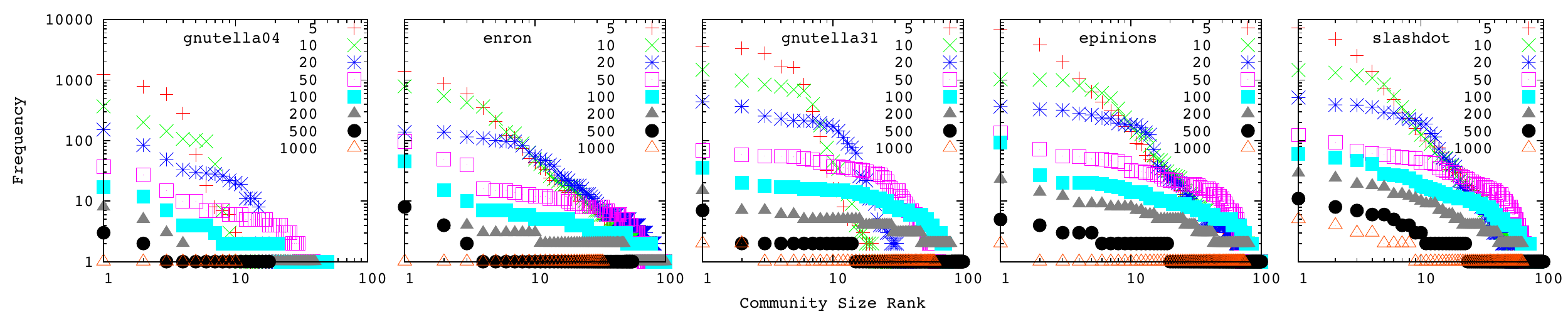}
  \caption{Distribution of the community size rank vs frequency observed in the different average size of communities and different real networks.}
  \label{fig:Comparison_Community_Sizes}
\end{figure*}

Figure~\ref{fig:Comparison_Community_Sizes} presents the rank distribution of the size of communities formed by this process, for the five networks studied and for various average community sizes, ranging from $5$ to $1000$ users/peer.
\tc{The distribution of the average community size for about $5$ users/peer exhibits power-law properties with one main exponent, especially for the social networks $enron$, $epinions$ and $slashdot$.
Merging smaller communities to form larger ones to increase the average community size up to $100$ users/peer leads to a cutoff in the power-law distribution of the community size, which can be calculated as shown in~\cite{clauset09powerlaw}.
Overall, the $Zipf$ distributions applicable in this range of community sizes shows that the communities formed maintain a power-law structure~\cite{clauset04community}.}
When the average community size is increased above $100$--$200$ users/peer, the $Zipf$ distribution is no longer applicable, as the communities become more uniform.

\begin{figure*}[htbp]
  \centering
	\includegraphics[scale=0.70]{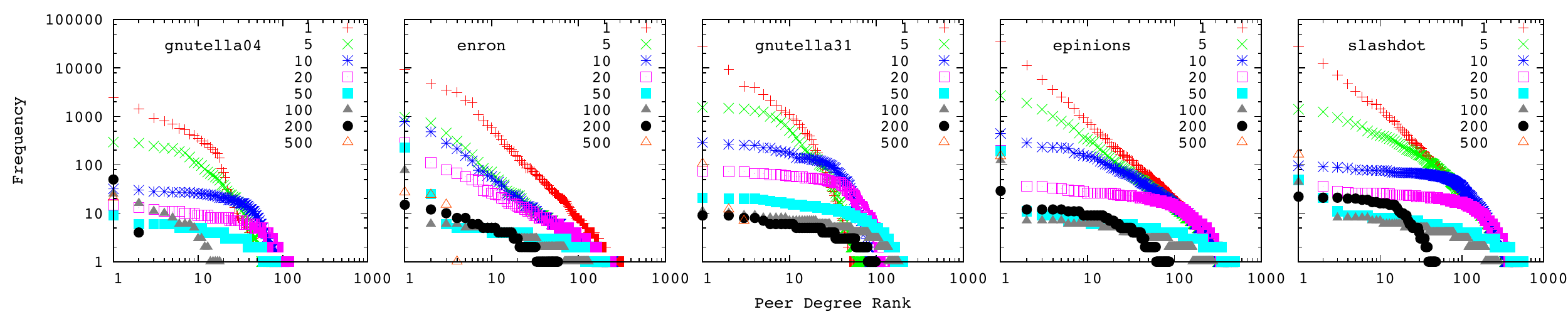}
  \caption{Distribution of the peer degree rank vs frequency observed in the different average size of communities and different real networks.}
  \label{fig:Comparison_Peer_Degrees}
\end{figure*}

Figure~\ref{fig:Comparison_Peer_Degrees} presents the rank distribution of the peer degree in the $PG$ for different average size of community, for each of the five networks examined.
The user degree rankings of the networks (points marked as ``1") follow a $Zipf$ distribution demonstrating a power-law nature (especially the larger networks \textit{epinions} and \textit{slashdot}).
Similar to the community size rankings, the networks exhibit a \tc{power-law distribution with a cut-off} when the size of communities increases from $5$ to about $20$--$50$ users per peer, meaning that the topologies inherit social structure from the $SG$ distributed on the peers.
Beyond a community size of about $100$ users, the topology becomes significantly uniform: most peers exhibit a similar degree, thus degree rankings show similar frequency.
This effect intensifies as the average community size increases to $1000$ users per peer.

\section{Centrality in Social and Projection Graphs}\label{sec:association-sg-pg}

In Section~\ref{sec:sna-measures} we ask if we can estimate the degree, node and edge betweenness centrality of peers and $PG$ edges when considering only local information, i.e., the cumulative scores of users (\tc{$SG$ edges}) mapped on peers ($PG$ edges).
In this section we examine experimentally how each of the three centrality metrics for a peer depends on the number of users mapped on the peer and their cumulative centrality metric, on the five real graphs and their extracted $PG$ topologies, as explained in Section~\ref{sec:pg-setup}.
In particular, we are interested to identify within what range of number of users per peer this estimation maintains high accuracy.

\subsection{Estimation of Centrality Measures}

\begin{figure*}[htbp]
  \centering
	\includegraphics[scale=0.7]{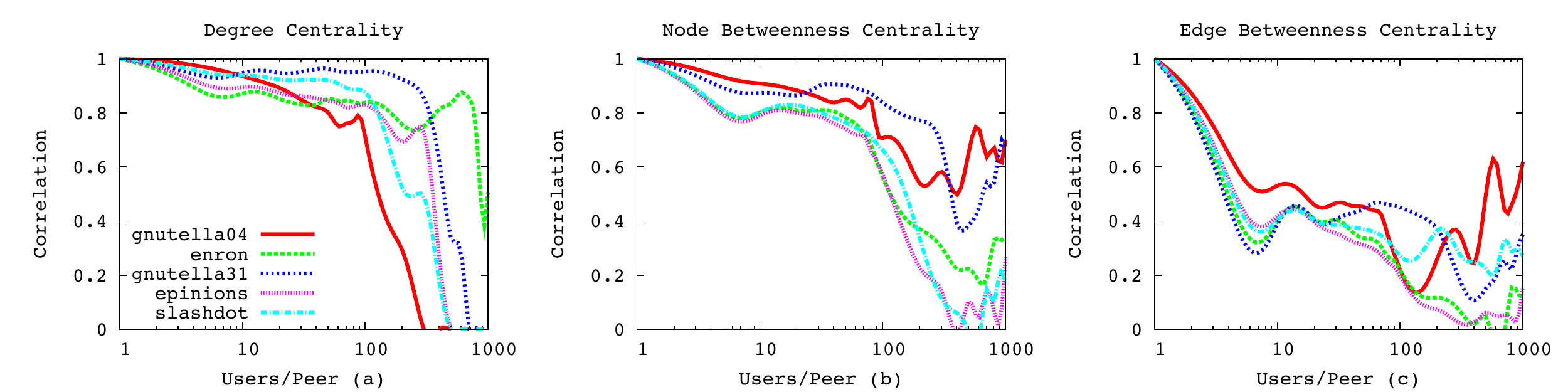}
  \caption{Correlation of cumulative normalized centrality scores of users vs normalized centrality scores of peers for Degree, Node Betweenness and Edge Betweenness Centrality.}
  \label{fig:Correlations_norm}
  \vspace{-3mm}
\end{figure*}

Figure~\ref{fig:Correlations_norm} presents for each metric the Pearson correlation of the scores of peers and cumulative scores of users per peer, with respect to the average number of users per peer \tc{(i.e., average community size)}.
Specifically, we calculate the correlation for each metric based on the tuple $\{A,B\}$ of scores per peer (or edge): (A) the cumulative centrality of users (or \tc{$SG$ edges}) mapped on peer $P_i$ (or $PG$ edge \tc{with $\Theta_{ij}$}), and (B) the centrality of the corresponding peer $P_i$ (or $PG$ edge \tc{with $\Theta_{ij}$}) in the resulting topology.

More precisely,
\[ \text{Degree Centrality: } \{ \sum_{u \in \Gamma_i} C_D(u), C_D(P_i) \}\]
i.e. (a) the cumulative degree of users mapped on a peer $P_i$, and (b) the degree of the corresponding peer $P_i$ in the resulting P2P network.
\[ \text{Node Betweenness Centrality: } \{ \sum_{u \in \Gamma_i} C_{NB}(u), C_{NB}(P_i) \}\]
i.e. (a) the cumulative node betweenness centrality of users mapped on a peer $P_i$, and (b) the node betweenness centrality of the corresponding peer $P_i$ in the resulting $PG$ topology.
\[ \text{Edge Betweenness Centrality: } \{ \sum_{e \in \Theta_{ij}} C_{EB}(e), C_{EB}((P_i,P_j)) \}\]
i.e. (a) the cumulative edge betweenness centrality of social edges mapped between peers $P_i$ and $P_j$, and (b) the edge betweenness centrality of the corresponding $PG$ edge $(P_i,P_j)$ in the resulting $PG$ topology.

The correlation is calculated by taking into account the tuples across all peers (or $PG$ edges) in the network, given a particular \tc{average community size}.

We observe that, for most of the networks, the correlation of the degree and node betweenness centrality \tc{between cumulative user centrality and peer centrality} remains fairly steady and high overall ($>0.8$) for communities of less than $100$-$200$ users.
From that point on, the correlation decreases rapidly.
This trend is generally consistent across all sizes and types of real graphs, but some networks present an outlier behavior.
For degree and node betweenness centrality, \textit{gnutella31} maintains a high correlation up to $300$ users/peer before the steep drop.
The edge betweenness centrality drops significantly when increasing the community size, demonstrating that it is more sensitive to this parameter than the degree or node betweenness centrality.
Next, we elaborate on the details behind the correlation performance of each centrality metric.

\begin{figure*}[htbp]
  \centering
	\includegraphics[scale=0.70]{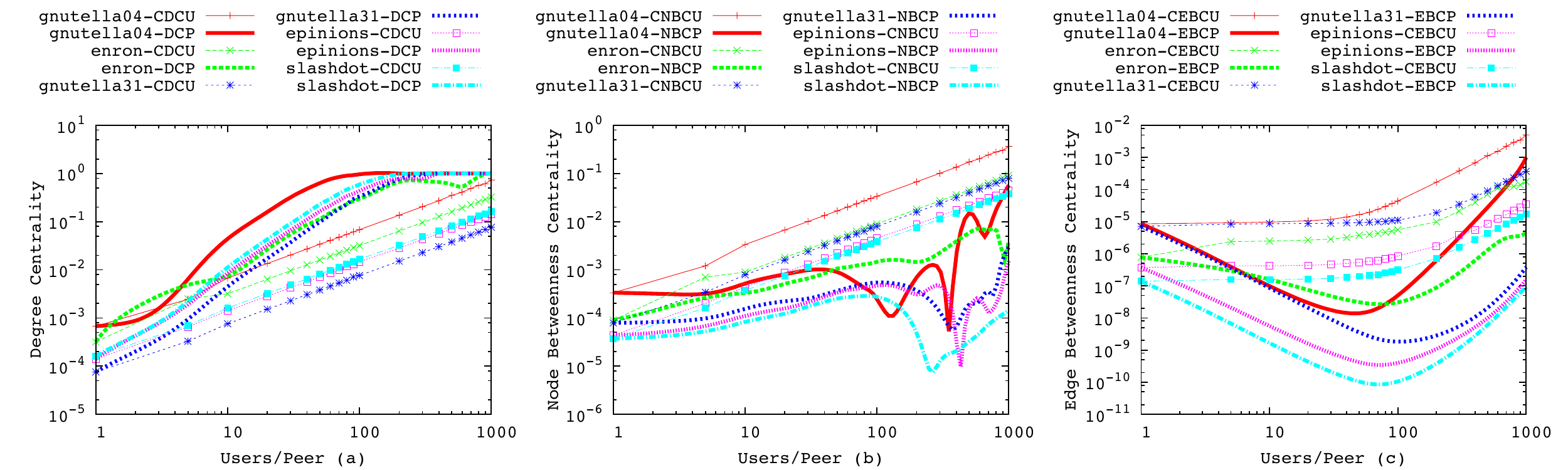}
  \caption{Comparison of cumulative normalized scores of users (point lines) vs average normalized scores of peers (smoothed lines) for Degree, Node Betweenness and Edge Betweenness Centrality.}
  \label{fig:Comparisons_norm}
  \vspace{-3mm}
\end{figure*}

\begin{figure}[htbp]
	\centering
	\includegraphics[scale=0.70]{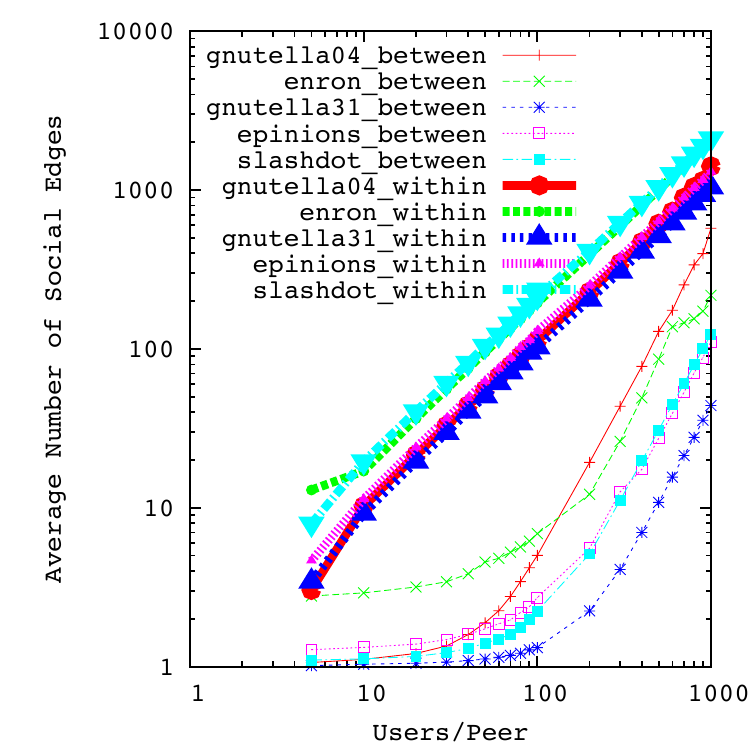}
	\caption{Average number of \tc{$SG$ edges} found within a peer (thick lines) or between two peers (thin lines).}
	\label{fig:SocialEdgesCounts}
\vspace{-3mm}
\end{figure}

Figure~\ref{fig:Comparisons_norm} compares the average degree centrality (DCP), node betweenness centrality (NBCP) and edge betweenness centrality (EBCP) for peers, with the respective cumulative centrality metric for users mapped on peers.
Figures~\ref{fig:Comparisons_norm}(a) and~\ref{fig:Comparisons_norm}(b) show \tc{an increase of} the degree and node betweenness centrality of peers \tc{when increasing the average community size}.
This means that adding more users to a peer directly \tc{increases} its centrality, according to these metrics.
We explain this as follows: by increasing the size of the communities we reduce their number, \tc{since we do not allow replication of user data in this experiment}.
In effect, more users mapped on a peer \tc{potentially means more $SG$ edges between these users and users of new peers}, thus more $PG$ edges to \tc{new} peers (i.e., higher degree centrality), as well as opportunity of the peer to participate in more \tc{PG} shortest paths (i.e., higher betweenness centrality).
Figure~\ref{fig:Comparisons_norm}(c) shows that the cumulative edge betweenness centrality of \tc{$SG$ edges} between peers does not change for a range of size of communities.
This is because when increasing the community size from $1$ to about $50$ users per peer, more \tc{$SG$ edges} are mapped \textit{within} peers instead of \textit{between} peers, as demonstrated in Figure~\ref{fig:SocialEdgesCounts}.
Within this range, the weighted $PG$ edge betweenness centrality decreases: the number of peers is reduced, and new \tc{$PG$ edges} between peers distribute the betweenness centrality of $PG$ edges across multiple \tc{$PG$} paths, thus losing importance (in terms of betweenness).

As the number of users mapped on the same peer increases, the degree and node betweenness centrality of peers reach a maximum point.
\tc{For the degree centrality this can be seen at the point where the graphs reach the plateau of 1 (the maximum possible value for the normalized degree centrality)} (e.g., the \textit{gnutella04} topology reaches a maximum average degree centrality at about $80$ users per peer.
From eq.~\ref{eq:dp}, when the average size of community per peer \tc{$\overline{|\Gamma_i|}$} increases, the second and third terms increase as well but at a \tc{higher} rate than the first term, thus the difference of them becomes least at this maximum point.
The network is optimally divided in communities mapped on peers which exhibit highest average degree and node betweenness centrality.
For the edge betweenness, this is a turning point: between $50$ and $100$ users per peer, more \tc{$SG$ edges} are mapped on $PG$ edges, in effect reversing the decline observed in Figure~\ref{fig:Comparisons_norm}(c).

Increasing further the average community size decreases rapidly the peer degree centrality to very small values (also verified by the flat distribution of peer degrees in Figure~\ref{fig:Comparison_Peer_Degrees}).
In addition, the opportunity to influence information flows (due to high \tc{peer} betweenness) is distributed uniformly across all peers since they start forming a small, tightly connected graph \tc{(as also seen by the plateau reached in the peer degree centrality in Figure~\ref{fig:Comparisons_norm}(a))}.
For the smallest network \textit{gnutella04}, this drop takes effect quickly at about $60$ users per peer, whereas for larger networks, like \textit{epinions} and \textit{slashdot}, at about $200$--$500$ users per peer.
At the same time, even though the betweenness centrality of $PG$ edges increases, the opportunity to influence information flows over indirect paths is distributed evenly across very few $PG$ edges.
Eventually, by increasing even further the community size, the peer degree reaches $0$, since at that point all users are mapped on one peer and this peer has no inter-peer edges.
It is important to note that depending on the application domain, the network properties of the topology may vary, even for seemingly small networks such as the \textit{enron} email graph in comparison to \textit{slashdot} or \textit{epinions} graphs.

Figure~\ref{fig:Comparisons_norm} helps explain the correlation performance in Figure~\ref{fig:Correlations_norm}.
Up to the \tc{turning} point for degree and betweenness centrality, the values of each pair of metrics increase with the addition of more users on each peer, and thus, the correlation is high overall.
After this point, there is rapid decrease in the centrality scores of peers but not for the cumulative scores of users, and this reverse relationship causes the steep drop in the correlation of the respective measures.
For the edge betweenness, the correlation drops early, as there is a high deviation between the $PG$ and $SG$ edge betweenness centrality scores (as explained earlier).

\subsection{Applicability of Results}
These experiments lead to the following lessons:\\
\textbf{Lesson 1: Community size vs. peer centrality.}
The increase of the average community size has an immediate effect on the $PG$ topology and thus on the social network measures of each peer.
We identified a turning point where the degree and node betweenness centralities of peers reach a maximum.
Before this point is reached, the $PG$ resembles more closely the $SG$ it projects.
Thus, the correlation of social network metrics between users and their peers is highest, and the measures for peers can be estimated with good accuracy by the cumulative scores of their users.
When this point is reached, the peers gain maximum opportunity to influence the information flows passing through them.
After this point, the topology loses any social properties, becomes a highly connected network and peers acquire equal opportunity to participate in $SG$ traversals.

\textbf{Lesson 2: Estimation of peer centrality.}
Users mapped on a peer reflect their $SG$ importance onto their peer in the P2P topology in two ways: either directly by connecting their peer with other peers (degree centrality), or indirectly by situating their peer on multiple shortest paths between other peers (betweenness centrality).
For small and medium size communities, we observe high correlation between users and peers for both of these centrality metrics.
Thus, the centrality scores of users acquired from local information available to peers are good predictors of the importance a peer will have in the network.
In effect, this means a peer can estimate with high accuracy its importance in the $PG$ without the need to traverse the P2P network, which might be difficult due to network size, peer churn and user data access policies on other peers.

\textbf{Lesson 3: Estimation of $PG$ edge centrality.}
There are high betweenness $SG$ edges that control significant information flow between different $SG$ parts.
\tc{However, the importance of a $SG$ edge reflected on the P2P network depends on how it is mapped in the topology.
We observed that when more $SG$ edges are mapped within instead of between peers, the estimation of $PG$ edge betweenness centrality from the cumulative centrality of $SG$ edges is less accurate.}

\section{Leveraging the Projection Graph}\label{sec:scenarios-application-overlay}

Our intuition is that the $PG$ centrality properties (either estimated as proposed earlier, or calculated exactly) can be used to improve the performance of applications such as social data search or dissemination of emergency messages.
In the following, we focus on social data search.
We study two different search workloads and investigate techniques that use the $PG$ model and its properties at the application level (Section~\ref{sec:social-graph-traversals}) and the overlay level (Section~\ref{sec:projection-graph-traversals}) for improved performance.

\subsection{Application Workloads}
Two applications inspired from the scenarios presented in Section~\ref{sec:motivating-scenarios} are described below. 
The first application is $person$-$finder$: the relatives of a missing person in the disaster search the civilian network distributed on the P2P network.
Social information is highly geo-localized in the system.
The search could start from a remote node close to the last known location of the missing person.
The search traverses the $SG$ to find time-relevant information about this person, either stored by him or by others.
Many community peers could be potentially visited before any information about the person is found.
A smart search technique should strive to limit the peers visited and thus the overall communication in the system, while maximizing the success rate.

The second application is $team$-$builder$ in online gaming, a service that builds teams by matching players based on their gaming characteristics such as play statistics or level of experience. 
Server administrators occasionally instantiate such a service for competitions or simply for increased fun.
Such a service aims to find $D$ players from at least $\sqrt D$ distinct communities (for diversity in playing style) in order to form $N$ teams with $C$ players each ($D$$\geq$$C$x$N$).
The service traverses the meta-gaming $SG$ in search for the right combination of players, potentially visiting hundreds of servers, with each server storing data of tens to hundreds of players.

These two applications represent the following search workloads in the $SG$:
a)~Starting from a random user \emph{s}, find a specific user \emph{d}. 
b)~Starting from a random user \emph{s}, find any \emph{D} users from $\sqrt D$ different communities. 

\subsection{Experimental Setup}

For these experiments we used the two largest graphs with their detected communities, i.e., $slashdot$ and $gnutella31$.
$Slashdot$ has about double the graph density than that of $gnutella31$, which leads to shorter average path lengths ($4.07$ vs. $5.94$ hops).
We constructed their $PG$ considering communities of about five users (a typical family server) with maximum sizes of $358$ and $97$ users for $slashdot$ and $gnutella31$ respectively (closer to gaming server size).
This average size also offers the best estimation of peer centrality from the cumulative scores of users, while allowing the formation of large-scale peer networks ($>10K$ peers).

\tc{We also validated our observations with real, user-declared mappings of users onto user-created groups on two large-scale networks, one for each of the application scenarios: the $LiveJournal$~\cite{mislove07measurementsOSNs} for $person$-$finder$ and the $Steam$~\cite{blackburn12games-cheaters} for the $team$-$builder$.
Next, we present the experimental setup using the $slashdot$ graph and discuss in more detail the two large-scale networks in Section~\ref{sec:case-studies}.}

$Person$-$finder$ search was performed for a number $S$ of different pairs of source and destination users.
This number was set to 10\% of the whole $SG$, i.e. $S$=$6256$ ($8216$) pairs for $gnutella31$ ($slashdot$).
$Team$-$builder$ search was performed for the same number of starting users $S$.
$D$ was set to $1\%$ of the users of each $SG$, to force search queries to traverse each $SG$ for more than 2 social hops (i.e., to visit at least the friends of friends of a source user): on average, for $\sim$$4$ ($\sim$$2.5$) hops for $gnutella31$ ($slashdot$), with user average degree $4.7$ ($13.2$).
We did not apply any constraints on the number of hops traveled from the source user, to study the highest possible success rate with respect to the incurred communication cost.
However, we maintained the history of the previously visited users/peers and stopped the search when either the search goal had been met, or all neighboring users/peers had been visited.
We measured the query success rate, the number of $SG$ and $PG$ hops traversed, and the percentage of system peers accessed (P2P communication overhead).

\subsection{Leveraging the Projection Graph at the Application Level}\label{sec:social-graph-traversals}

In the first approach, we inform the search not only with $SG$ topology properties~\cite{adamic01search}, but also with $PG$ properties which peers acquire within the system. 
The intuition is that a search query should be forwarded to users who being mapped on central peers are likely to be connected to other central users mapped on the same central peers~\cite{shi08well-conn}, and this should lead to improved search performance.
In this section we investigate how the $PG$ properties can inform various social search techniques to increase the success rate and reduce overhead.

\subsubsection{Social Search Techniques}\label{soc-search-tech-app-level}

We investigate the following four search techniques for traversing the $SG$.
The first three techniques assume that during the $SG$ traversal, a user forwards the search query to its neighboring users mapped on peers with 1)~degree centrality in the top$N\%$ of neighboring peers, 2)~betweenness centrality in the top $N\%$ of neighboring peers, or 3)~to neighboring users whose peers connect over $PG$ edges with betweenness centrality in the top $N\%$ of neighbor $PG$ edges.
These techniques allow an application to utilize peer centrality to inform its graph traversal when specific user centrality is not available (e.g., due to privacy settings).
We compare their performance with a baseline technique (4) which even though still utilizes the same $SG$ topology, it does not take into account $PG$ topology properties but randomly selects the same number $N$ of neighboring users to forward the query.
We tested these techniques for $N$=$20$ and $N$=$50$.

\subsubsection{Experimental Results}

\begin{figure*}[htbp]
	\centering
	\subfigure[CDF of the number of $SG$ hops for successful queries]{
		\includegraphics[scale=0.7]{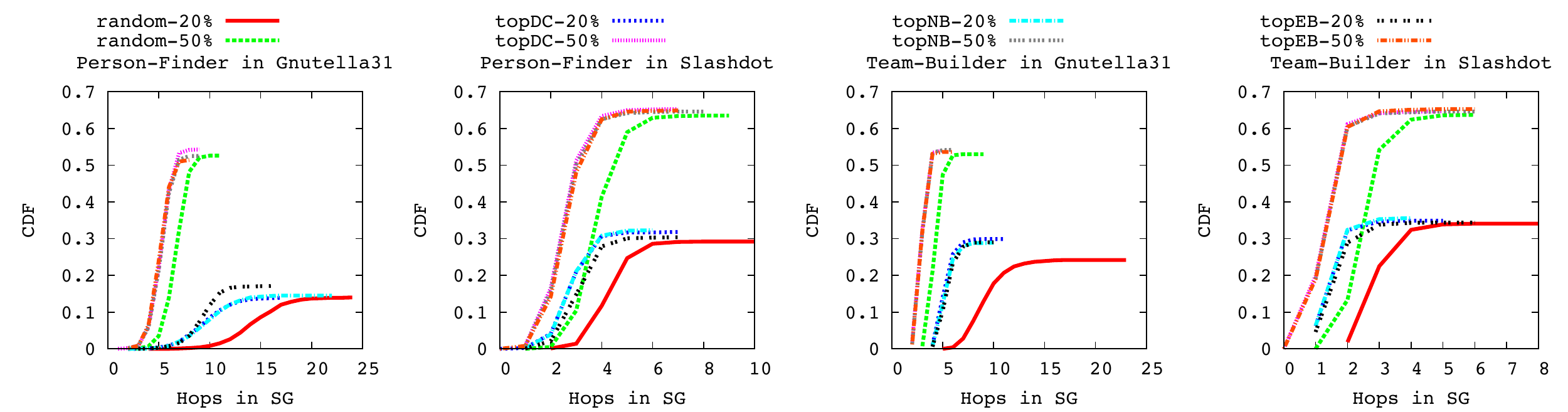}
		\label{fig:sg-st1-st2-hops-onlypeers}
	}
	\subfigure[CDF of the percentage of peers accessed in the system]{ 
		\includegraphics[scale=0.7]{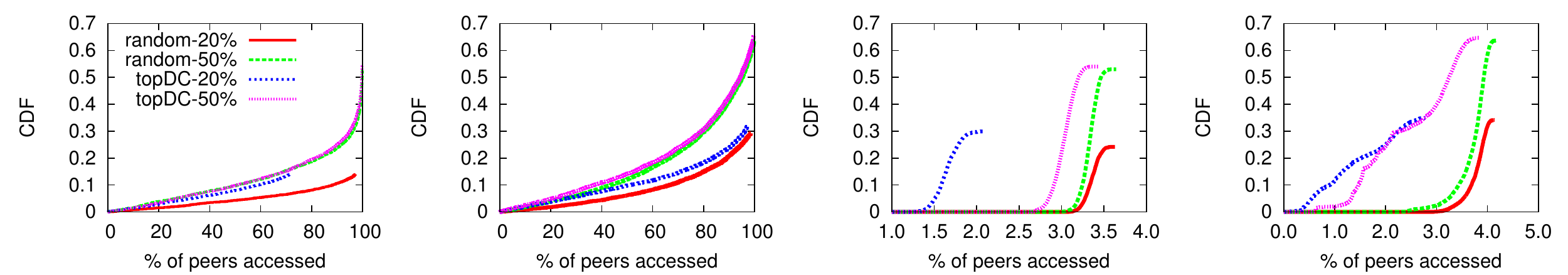}
		\label{fig:sg-st1-st2-overhead-peers-onlypeers}
	}
	\caption{\tc{Number of hops and system overhead of successful $person$-$finder$ and $team$-$builder$ queries in the $SG$, for different search techniques and portion of $SG$ edges used, for the networks $gnutella31$ (62,561 nodes and 147,878 edges) and $slashdot$ (82,168 nodes and 504,230 edges).}}
\vspace{-3mm}
\end{figure*}

\tc{Figure~\ref{fig:sg-st1-st2-hops-onlypeers} presents for the successful queries of the $person$-$finder$ and $team$-$builder$ searches, the CDF of the number of $SG$ hops traversed for the four search techniques.}
For the $person$-$finder$ search over $gnutella31$ ($slashdot$), all techniques converge to a maximum of about $55\%$ ($65\%$) of query success rate when $50\%$ of \tc{$SG$ edges} are used and about $17\%$ ($30\%$) when $20\%$ of \tc{$SG$ edges} are used.
For $gnutella31$ ($slashdot$), we notice that more than $50\%$ of the queries finish within $7$ ($3$) hops when centrality techniques are used in comparison to about $9$ ($4.5$) hops for the random technique.
Thus, even though the random technique uses the same number of edges on the same graph as the centrality techniques, the random selection of which edges to follow in the search leads to longer walks on the $SG$ and lower success rates.

\tc{Note that this application scenario ($person$-$finder$) is highly geo-localized, i.e., the search for a particular person follows the geographically distributed $SG$ edges or time-related ties between persons connecting the source and destination users.
Thus, while traditional DHTs are better at finding ``the needle in the haystack'', they are impractical in this scenario for the following reasons.}
First, identical names can exist within the same geographical region (and even within the same family) and thus ambiguity can be introduced as to which person's social data were returned.
Second, DHTs do not easily exploit the geographic locality implicit to this search type, which has to follow the \tc{$SG$ edges} within each community and geographic location to find the correct person from the appropriate community.

In comparison to the $person$-$finder$ search, the overall success rates reported for $team$-$builder$ search are $5$--$10\%$ higher, with the queries finishing within shorter walks.
This is expected as the $team$-$builder$ search is satisfied with \emph{any} users discovered in the social neighborhood of the randomly selected source, given they come from $\sqrt N$ different communities, as opposed to finding a specific user.
We notice that within $4$ ($2$) $SG$ hops in $gnutella31$ ($slashdot$), more than $50\%$ ($60\%$) of the queries finish using centrality techniques, whereas only about $20\%$ ($15\%$) using the random technique.

Figure~\ref{fig:sg-st1-st2-overhead-peers-onlypeers} shows the overhead as the percentage of peers accessed.
We compare the random technique only with the $topN\%$ peer degree centrality technique as the other centrality techniques perform similarly in success rate.
The $person$-$finder$ search on both networks has a similar overhead in both types of techniques, especially when using $50\%$ of \tc{$SG$ edges}.
However, the $team$-$builder$ search with the centrality technique has $0.25$--$2$ times less overhead than the random technique.

\tc{We also compared these results (i.e., using $PG$ properties), with techniques that used the centrality properties of the $SG$ (not shown here for brevity)
and observed that both $PG$ and $SG$ centrality techniques perform very similarly, and especially when using degree and node betweenness.}
This can be attributed to the high correlation between user and peer scores for the same centrality metric.
Furthermore, global metrics that require knowledge of the whole $SG$ or $PG$, such as node betweenness, do not add much gain in the search performance, so an application can effectively use local information instead, such as degree centrality.

\subsection{Leveraging the Projection Graph at the P2P Overlay Level}\label{sec:projection-graph-traversals}

Leveraging $SG$ knowledge has been applied to both structured~\cite{marti06sprout} and unstructured~\cite{popescu04turtle} P2P overlays. 
In this section we investigate the benefits of informing P2P overlay design and routing decisions in the system with $PG$-specific information.
We focus on unstructured overlays, leaving the structured overlays for future work.

\subsubsection{$PG$-Based Unstructured P2P Overlays}

By definition, a $PG$ is the accurate representation of the $SG$ mapped on the P2P system. 
We propose an unstructured P2P overlay that exactly mimics the $PG$: the routing tables in the P2P network consist of (a subset of) the $PG$ edges that connect different peers.
This overlay reflects well the social relationships between users (thus best supporting socially-aware applications), and, implicitly encapsulates geographical (and network) locality and clustering, since social relationships are usually geographically close~\cite{scellato11LBSN-distance}.
However, the power-law nature of the node degree in the $PG$ translates into high-degree peers maintaining unrealistically many connections.
Therefore, we propose that the $PG$ edges ($E_P$) are considered as \emph{potential} communication connections between peers in the overlay, but only some of them are implemented into \emph{active} communication connections ($E_A$), i.e., $E_A\subseteq E_P$.

We investigate the same four techniques but this time in overlay routing.
In the first three techniques, a peer forwards the search query to its neighboring peers with 1)~degree centrality in the $topN\%$ of the neighboring peers (set $D$), 2)~betweenness centrality in the $topN\%$ of the neighboring peers (set $B$), or 3)~that are connected over $PG$ edges with betweenness centrality  in the $topN\%$ of the neighbor $PG$ edges (set $E$).
The fourth technique is for baseline comparison: a peer forwards the query in the $PG$ topology to the same number $N$ of randomly selected neighboring peers (set $R$, $|R|$=$|D|$).
The difference from the application-level techniques (Section~\ref{soc-search-tech-app-level}) is that a query traverses the $PG$ instead of the $SG$.
Thus, instead of forwarding a message along $SG$ edges and potentially bouncing multiple times between the same peers, the message is forwarded along the $PG$ edges, thus reducing redundant communication.

To apply these techniques, the following assumptions are made:
First, a query can access all user data stored on a peer.
Second, as in all P2P systems, peers regularly update information regarding peers in their routing table, such as availability, but also $PG$-based centrality metrics.
Third, peers rank their neighbors based on the $PG$ metrics, and depending on the heuristic applied, they select the $topN\%$ subset as their active connections.
Therefore, depending on the search technique used ($t$=$1$,\dots,$4$), an active connection between $P_i$ and $P_j$ included in the set of active connections of $P_i$ (set $E_A^{P_i}$) is defined as follows:
\[ (P_i, P_j) \in E_A^{P_i} \text{  iff  } \exists P_i \in V_P, \exists P_j \in V_P \text{ s.t. } \]
\begin{equation}
(P_i, P_j) \in E_P \text{ and }
\begin{cases}
P_j \in D		&	\text{ if } t=1	\\
P_j \in B		&	\text{ if } t=2	\\
(P_i, P_j) \in E	&	\text{ if } t=3	\\
P_j \in R		&	\text{ if } t=4
\end{cases}
\end{equation}
Consequently, the total set of active connections in the P2P network $E_A$ is the union of the sets $E_A^{P_i}$ for all peers:
\begin{equation}
E_A = \bigcup_{\forall P_i \in V_P} E_A^{P_i}
\end{equation}

\subsubsection{Experimental Results}

\begin{figure*}[htbp]
	\centering
	\subfigure[CDF of the number of $PG$ hops for successful queries]{
		\includegraphics[scale=0.7]{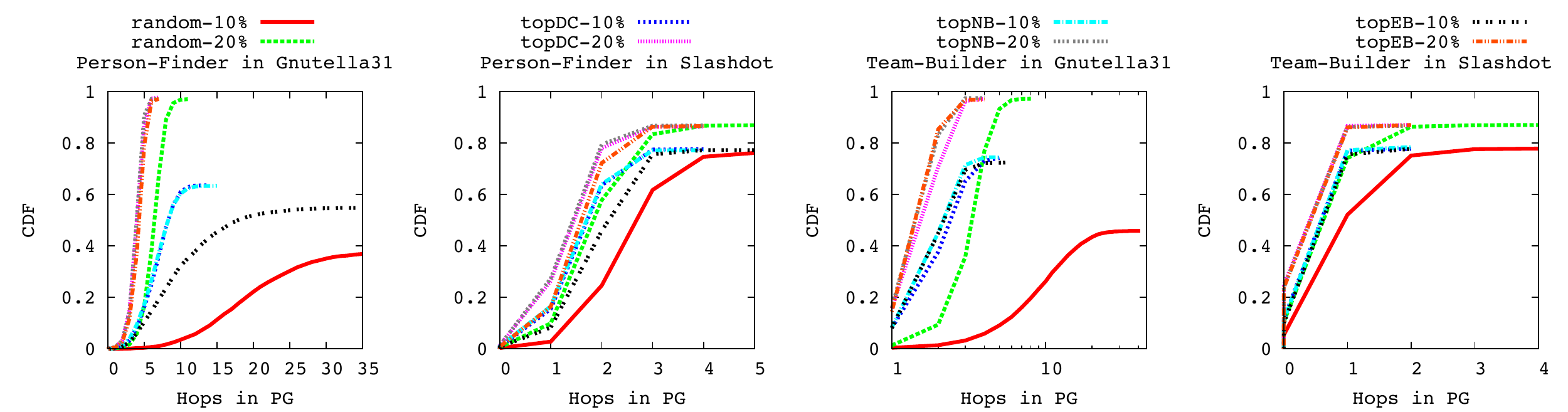}
		\label{fig:pg-st1-st2-hops}
	}
	\subfigure[CDF of the percentage of peers accessed in the system]{ 
		\includegraphics[scale=0.7]{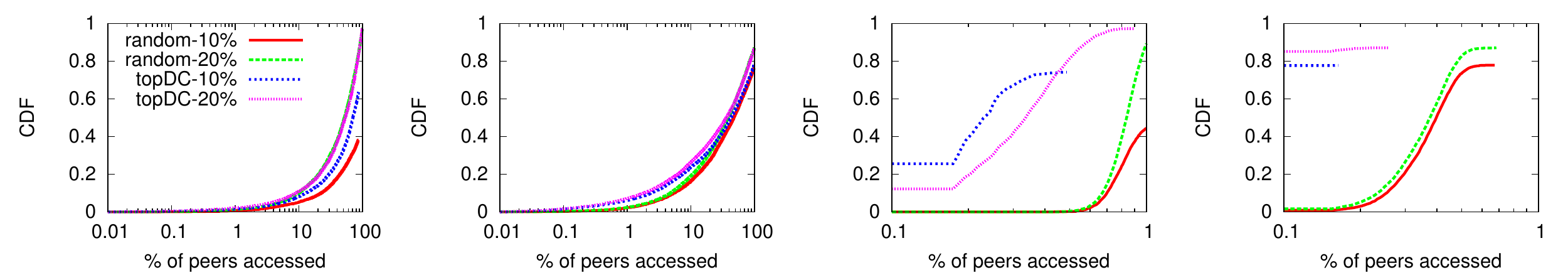}
		\label{fig:pg-st1-st2-overhead-peers}
	}
	\caption{\tc{Number of hops and system overhead of successful $person$-$finder$ and $team$-$builder$ queries in the $PG$, for different search techniques and portion of $PG$ edges used, for the networks $gnutella31$ (62,561 nodes and 147,878 edges) and $slashdot$ (82,168 nodes and 504,230 edges).}}
\vspace{-3mm}
\end{figure*}

We tested the four techniques by varying $N\%$, the portion of \tc{$PG$ edges} used.
Since the search query has access to all the users' data stored on a particular peer, we expected the search to finish with higher success rate and in shorter walks than when the search traversed the $SG$ edges (Section~\ref{sec:social-graph-traversals}).
By varying the portion of \tc{$PG$ edges} used from $1\%$ to $50\%$, our experiments (not shown here for brevity) revealed that using $\sim$$20\%$ of available $PG$ edges leads to almost maximum success rate for both $person$-$finder$ and $team$-$builder$ searches; above $20\%$ there is mostly increase in the message overhead with minor gains in success rate.
Using below $10\%$ of $PG$ edges leads to low search performance for both search types regardless of the technique used, but with the random technique performing the poorest.
Next, we compare the techniques using $10\%$ and $20\%$ of $PG$ edges.

\tc{Figure~\ref{fig:pg-st1-st2-hops} presents for the successful queries of the $person$-$finder$ and $team$-$builder$ searches, the CDF of the number of $PG$ hops.}
For the $person$-$finder$ search and $gnutella31$ ($slashdot$), all techniques converge to a maximum of about $98\%$ ($87\%$) of query success rate when $20\%$ of $PG$ edges are used and about $64\%$ ($78\%$) when $10\%$ of $PG$ edges are used.
A centrality technique on $slashdot$ using $20\%$ $PG$ edges has $20\%$ more success within $2$ P2P hops than the random technique.
This difference is amplified for $gnutella31$, where within $5$ hops the centrality techniques achieve over $60\%$ more success than the random technique.
For the $team$-$builder$ search, similar maximum success rates with the $person$-$finder$ search are reported for both networks, but with the queries finishing in shorter walks by $3$ ($2$) fewer P2P hops for $gnutella31$ ($slashdot$).

The gain in success rate and with fewer P2P hops is reflected on the system overhead presented in Figure~\ref{fig:pg-st1-st2-overhead-peers} as the percentage of peers accessed.
We compare the random technique only with the $topN\%$ peer degree centrality technique, as the other centrality techniques perform similarly in success rate.
Overall, the centrality technique leads to similar or lower system overhead than the random technique.
The $person$-$finder$ search has a similar overhead in both techniques, especially when $20\%$ of \tc{$PG$ edges} are used: up to $76\%$ ($99.4\%$) of peers were accessed in $gnutella31$ ($slashdot$) to reach maximum possible success rate.
For the $team$-$builder$ search, the random technique has about $3$--$4$ times more overhead than the centrality technique, due to longer walks in the $PG$ (as seen in Figure~\ref{fig:pg-st1-st2-hops}).
Thus, the technique needs to access a larger portion of peers to satisfy the $team$-$builder$ queries.

\subsection{Case Studies: Steam and LiveJournal}\label{sec:case-studies}

In order to validate our lessons from the previous experiments, we used traces from two large real networks for which we had real group membership information: Steam Community (\texttt{http://www.steamcommunity.com}) and LiveJournal (\texttt{http://www.livejournal.com}).

The Steam Community is a social network used by millions of online gamers around the world to declare friendships, organize online game sessions, exchange ideas and comments on the games, participate in discussion groups, etc.
Using the dataset from~\cite{blackburn12games-cheaters}, we constructed a social network of $12.5$ million players with $88.6$ million friendship edges.
From these, $5.3$ million players participate in $1.5$ million user groups spanning various topics.
On average, each user joined $3$--$4$ user groups and the groups have an average size of $21$ players.
LiveJournal is a large-scale blogging site whose users form a social network.
Using the dataset from~\cite{mislove07measurementsOSNs}, we constructed a social network of $5.2$ million bloggers with $77.4$ million friendship edges.
Using declared interests in the online profiles of $3.2$ million bloggers, $7.5$ million groups were formed.
On average, each blogger was mapped to $21$ groups, and the groups have an average size of $15$ bloggers.

In general, individuals join user-created groups due to interest in the group topic and its members, even if they are not directly connected with them over the social network with friendship edges.
This type of groups represent tightly clustered communities of users in the social network, as demonstrated by the high average clustering coefficient of group members~\cite{mislove07measurementsOSNs}.
Therefore, such groups comprise real communities with common objectives and real incentives for resource sharing~\cite{antoniadis07communities} and lead to a realistic mapping of users onto peers. 
We make the same assumption as before: each group is mapped on a member-contributed peer.

In these traces, users belong to multiple groups whereas our previous experiments considered that a user belongs to only one community. 
In order to be able to compare results, we selected the most representative group for each user, ignoring his membership to other groups. 
To this end, if $Bob$ participated in more than one group, we ranked each of his groups based on the Jaccard similarity coefficient $J(F,G)$ of his 1-hop friendship neighborhood (set $F$) and the group members (set $G$): $J(F,G) =| F \cap G | / | F \cup G | $.
Then, we mapped him to his highest ranking group, as it indicates the group that $Bob$ has the most social affinity and thus incentives to participate and contribute.
Table~\ref{tab:steam-livejournal} summarizes the sizes of groups and $SG$s before and after filtering.
Using these groups and the $SG$ edges between group users, we constructed a $PG$ for each network: 1)~Steam $PG$ with $848,519$ peers and $24,685,977$ $PG$ edges, 2)~LiveJournal $PG$ with $786,312$ peers and $21,171,246$ $PG$ edges.

\begin{table}[htdp]
\caption{Steam and LiveJournal groups and social graphs before and after filtering.}
\begin{center}
\begin{tabular}{|c|l|l|}
\hline
									&	\textbf{Steam Community}				&	\textbf{LiveJournal}						\\ \cline{2-3}
\multirow{3}{*}{\rotatebox{90}{\textbf{Before}}}	&	Groups:	1487551						&	Groups:	7489073						\\ \cline{2-3}
									&	Average Group Size:	20.94			&	Average Group Size:	15				\\ \cline{2-3}
									&	SG:		V=5227911 (E=46120284)		&	SG:		V=3067765 (E=29627072)		\\ \hline \hline
\multirow{3}{*}{\rotatebox{90}{\textbf{After}}}	&	Groups:	848519						&	Groups:	786312 						\\ \cline{2-3}
									&	Average Group Size:	6.14				&	Average Group Size:	3.87				\\ \cline{2-3}
									&	SG:		V=5208315 (E=46105938)		&	SG:		V=3042072 (E=29597844)		\\ \hline
\end{tabular}
\end{center}
\label{tab:steam-livejournal}
\end{table}

\begin{figure*}[htbp]
\begin{center}
	\includegraphics[scale=0.8]{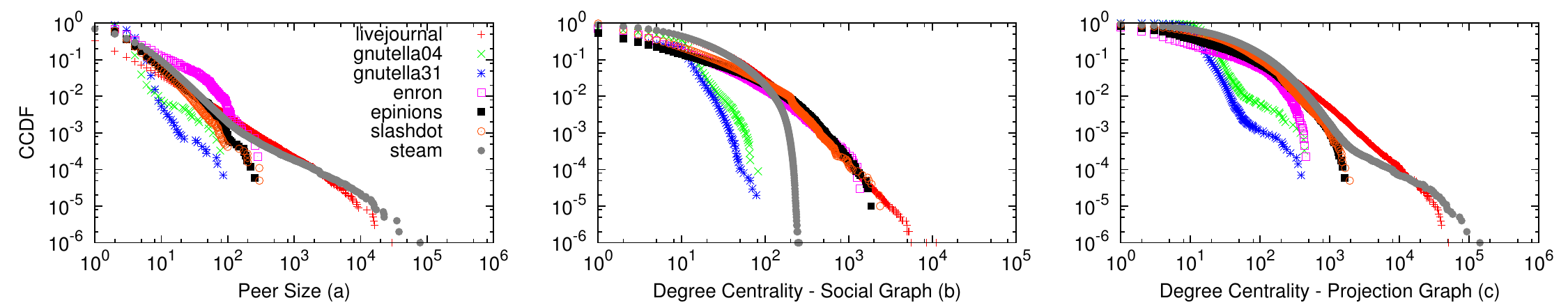}
\caption{CCDF of the (a)~peer sizes, (b)~the user degree centrality and (c)~the peer degree centrality, for the Steam and LiveJournal social networks.
The other 5 networks (for community size of 5 users/peer) are plotted as well, for comparison.}
\label{fig:dc-sizes}
\end{center}
\end{figure*}

Figure~\ref{fig:dc-sizes}(a) demonstrates the power-law nature of the distribution of group sizes declared by users and Figures~\ref{fig:dc-sizes}(b) and~\ref{fig:dc-sizes}(c) show the distribution of the user and peer degree centrality, respectively, for these two large social networks and how it compares with the other real networks presented earlier.
The distribution of user degree centrality for $enron$, $epinions$, $slashdot$ and $livejournal$ is almost identical, which verifies the social nature of all these networks and that our results in the previous section are applicable in social networks regardless of size.
The degree centrality distribution of $steam$ users deviates from that of the other networks, due to the company-imposed upper limit of 250 friends per player (as seen by the long tail).
However, the degree centrality of peers in the $PG$ follows closely the distribution of the LiveJournal and the other $PG$s.

To compare with the previous correlation results, we also calculate the correlation of the cumulative degree centrality of users in the $SG$ with respect to the degree centrality of their peers in the $PG$. 
We find $\rho$=$0.964$ for the $Steam$ and $\rho$=$0.946$ for the $Livejournal$ network, which confirm our previous results that the degree centrality of users and their peers is highly correlated, especially in this range of users per peer.

\begin{figure*}[htbp]
	\centering
	\subfigure[CDF of the number of $SG$ or $PG$ hops for successful queries]{
		\includegraphics[scale=0.7]{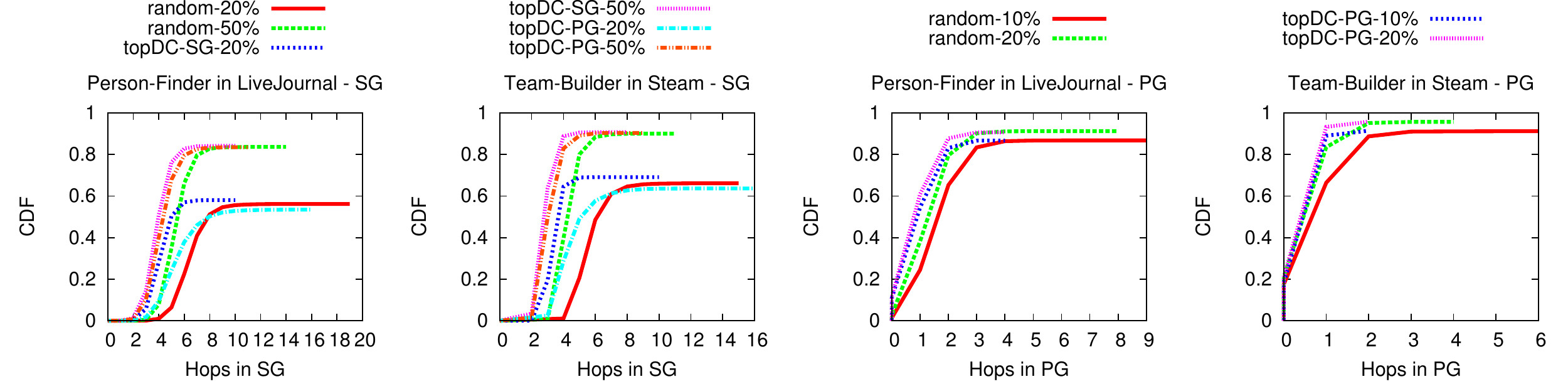}
		\label{fig:ST1-ST2-CDF-hops-large-nets}
	}
	\subfigure[CDF of the percentage of peers accessed in the system]{ 
		\includegraphics[scale=0.7]{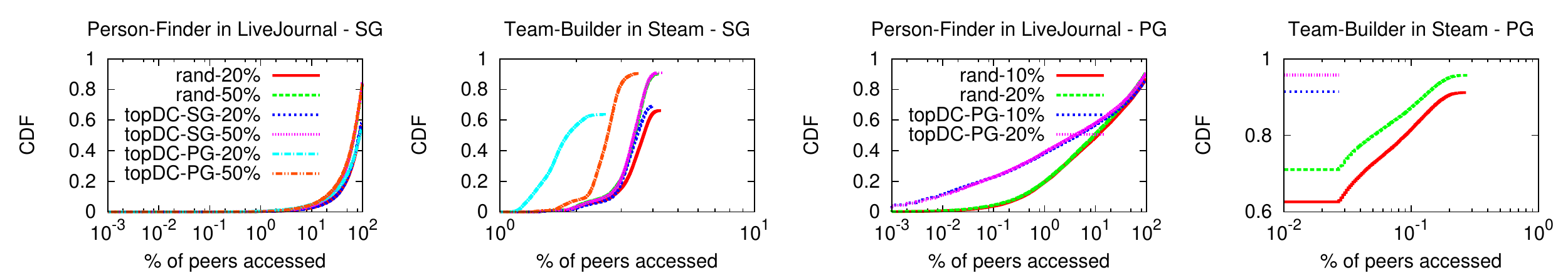}
		\label{fig:ST1-ST2-CDF-overhead-peers-large-nets}
	}
	\caption{Success rate and system overhead of successful $person$-$finder$ and $team$-$builder$ queries in the $SG$ and $PG$, for different search techniques and portion of edges used, for the $Livejournal$ and $Steam$ networks.}
\vspace{-3mm}
\end{figure*}

Figures~\ref{fig:ST1-ST2-CDF-hops-large-nets} and~\ref{fig:ST1-ST2-CDF-overhead-peers-large-nets} show the performance of the random and degree centrality techniques for the $person$-$finder$ search on the $LiveJournal$ $SG$ and $PG$, and for the $team$-$builder$ search on the $Steam$ $SG$ and $PG$ using the previously described groupings.
These results confirm our observations on $slashdot$ with detected communities.
Using the degree centrality of users or peers, the centrality techniques lead to increased search success rate within fewer $SG$ or $PG$ hops than a random technique.
Consequently, shorter walks in the graph lead to reduced system overhead, as expressed by the percentage of peers accessed during the search (Figure~\ref{fig:ST1-ST2-CDF-overhead-peers-large-nets}).
Furthermore, in some cases the $PG$ degree centrality is a better metric to use when making forwarding decisions, instead of $SG$ degree centrality.
This is because the $PG$ metric summarizes the centrality of a group of users and can help direct faster the search to more central parts of the graph.

\subsection{Applicability of Results}

\tc{The experimental results from search techniques on different social graphs in both the application and overlay layer provide support to our initial intuition. $PG$ centrality properties can be used to improve social search performance, i.e., reduce the number of graph hops and increase success rate, while reducing system overhead.
In particular, we formulate the following lessons:}

\textbf{Lesson 4: $SG$ vs. $PG$ traversals.}
In comparison to traversing the $SG$ edges, leveraging the $PG$ topology provides access to social information of more users and thus, on average, increases the success rate by $10\%$--$25\%$, reduces the walk length by $1$--$2$ hops and decreases the percentage of peers accessed by $40\%$ for the $person$-$finder$ search and by $2.5\%$ for the $team$-$builder$ search.
Thus, socially-aware applications and services could be developed to take full advantage of the available information for enhanced application search and overlay performance.

\textbf{Lesson 5: Centrality vs. Random Techniques.}
The centrality techniques lead to higher success rates within fewer hops (in $SG$ or $PG$) than the random technique.
In particular, even though the random technique is also socially-aware as it utilizes the same $SG$ and $PG$ topology construction as the centrality techniques (but randomly selects to which users or peers to send the query), it still requires about $1$--$3$ more hops to reach the same success rate as the centrality techniques, thus imposing higher overhead in the system.

\textbf{Lesson 6: User vs. Peer Centrality.}
Search techniques that use $SG$ or $PG$ centralities perform similarly.
This means that an application could select which of the centrality techniques to use based on the available centrality information for each user or peer.
If it cannot access the individual score of Alice in the $SG$ or her full data to calculate it (e.g., due to privacy settings), but can access her peer's score (an aggregate metric for a user group), it can achieve the same performance by routing queries through the $SG$ edges using the $PG$ peer centralities.

\textbf{Lesson 7: Local vs. Global Information.}
Search techniques that use global centralities calculated over the whole graph (i.e., node and edge betweenness centrality) perform similarly to the ones calculated using local information (i.e., degree centrality).
Therefore, an application can utilize the degree centrality of users or peers to inform the forwarding decision of the search query.

\textbf{Lesson 8: $PG$-based Overlays.}
A P2P system can leverage a centrality technique that uses local information such as peer degree centrality to construct the set of active connections $E_A$ used by peers during a search.
Furthermore, a small set of active connections $E_A^{P_i}$ per peer is enough to ensure high performance and low communication overhead.
This fraction of \tc{$PG$ edges} would mean for the most connected peer of $gnutella31$ and $slashdot$ a maximum of $80$ and $392$ active connections respectively, which is well below the maximum connections of deployed unstructured overlays (Gnutella V0.6 had peers with more than $500$ connections~\cite{rasti06gnutellav06}).

\section{Summary and Discussions}\label{sec:discussion}

In this work we proposed the \emph{projection graph} ($PG$), a model for studying the network properties of a P2P system that hosts the social graph ($SG$) of its users in a distributed fashion.
We represented analytically the degree, node and edge betweenness centrality for the $SG$ and $PG$ and discussed the relation between the two types of graphs in terms of these centrality measures.
Because the analytical expressions are heavily dependent on the topology of these graphs and how the $SG$ is distributed on the P2P system to form the $PG$, we studied experimentally the correlation between their centrality metrics over a wide range of configurations.
Our experiments showed that within a range of $50$--$150$ users mapped on a peer, there is an optimal organization of the $PG$, since peers score the highest average degree and node betweenness centrality.
Also, up to this point, there is a high correlation between the properties of users and their peers, which degrades rapidly when the number of users per peer passes this threshold.
This correlation allows us to estimate with high accuracy the centrality of peers based on the centrality scores of users.

In addition, we investigated experimentally \tc{in large-scale social graphs how $PG$} peer centrality metrics can be used for the $SG$ traversal of search applications.
We found that targeting top ranked degree peers in the $PG$ or top ranked degree users in the $SG$ can achieve equal improvements in the performance of a social search.
This is true when the search is executed either at the application or overlay layer.
\tc{Thus, if an application was not granted access to individual user centrality scores in the $SG$ (due to user privacy settings), but only to peer scores in $PG$, it can still route search queries through the $SG$ or $PG$, using the peer degree centralities.}
These results are applicable to other work~\cite{gummadi06exploiting,yang09p2psearch,lin07socialunstructured}, where social search can be informed using an estimation of the $PG$ peer degree centrality.

\tc{In our experiments, we increased the average community size from $5$ to $1000$ users/peer, which inevitably decreased the number of peers in the system (from thousands to tens).
This parameter allowed} us to study how the degree of social data decentralization affects the network properties of the nodes in the system.
Large-scale systems such as mobile phone or P2P networks decentralize the users' social data on thousands of nodes and allow each device to access social data of a small set of users.
We observed that the smaller this set, the higher the association is between the users' centrality in the $SG$ with their device's centrality in the $PG$, and an application can use either centrality score (user or peer) to effectively route search queries.
On the other hand, centralized company systems with a few hundred machines enable each node to access social data of thousands of users.
By distributing the social data on centralized machines in a socially-aware manner (e.g., as in~\cite{pujol10little-engines}), our experiments reveal that the degree and node betweenness centrality of peers should be similar, and thus all peers should have an equal opportunity to be queried for social data.

\tc{We demonstrated experimentally on large-scale social networks, with user-created groups and user-declared participation to groups,} the benefits of building an unstructured P2P overlay by leveraging the $PG$ topology and selecting P2P overlay links using centrality metrics.
\tc{Our results on social search show that overlay overhead can be reduced if peers construct their routing tables using $PG$ edges to neighboring central peers.}
However, such P2P network paths can be used frequently from any type of application traversing the $SG$ and not only from social search.
\tc{Thus, these paths should be explicitly defined and used in the P2P overlay construction.}

This way of overlay construction could be embedded in systems already implementing a socially-informed design (e.g., Prometheus~\cite{kourtellis10prometheus}, Turtle~\cite{popescu04turtle} and Sprout~\cite{marti06sprout}), but instead of using \textit{single} \tc{$SG$ edges between users, they could exploit high weight $PG$ edges which represent \textit{multiple} $SG$ edges between groups of users, and indicate stronger social ties and potential trust between users, and consequently, their peers.
These peer paths lead to more secure discovery of new peers for data hosting within reduced network hops.}
Moreover, such $PG$ edges could represent social incentives between multiple users for data sharing among neighboring communities and their peers.
\tc{Thus}, potential increase of the communication between these peers when serving application workload, or for system maintenance due to peer churn, can be tackled with data replication to neighboring peers in the $PG$ for better data availability and load distribution.

Taking it a step further and using intuition from sociological studies, a P2P system could predict the creation of social edges between users, by monitoring the triadic closures between them and identifying which ones violate the \emph{forbidden triad rule}~\cite{granovetter73weakties}.
This rule refers to the situation where two individuals, not socially connected with each other but with a strong social connection with another mutual individual, will likely form a social connection with each other in the future.
This observation could enable the system to anticipate access of the particular users' social data, and thus perform proactive caching on central or neighboring peers.

\section*{Acknowledgment}
This research was supported by the National Science Foundation under Grants No. CNS 0952420 and CNS 0831785.
Any opinions, findings, and conclusions or recommendations expressed in this material are those of the authors and do not necessarily reflect the views of the sponsors.

\bibliographystyle{IEEEtran}

\bibliography{refs}

\end{document}